\documentclass[conference]{IEEEtran}
\IEEEoverridecommandlockouts
\usepackage{algorithm}
\usepackage[noend]{algcompatible}
\usepackage{amsmath,amssymb,amsfonts}
\usepackage{booktabs}
\usepackage{cite}
\usepackage{enumitem}
\usepackage{graphicx}
\usepackage{multirow}
\usepackage{subcaption}
\usepackage{textcomp}
\usepackage{xcolor}
\usepackage{url}
\newcommand{\T}{\mathcal{T}}

\def\BibTeX{{\rm B\kern-.05em{\sc i\kern-.025em b}\kern-.08em
    T\kern-.1667em\lower.7ex\hbox{E}\kern-.125emX}}
    
\begin{document}

\title{\title{Fast Algorithms for Scheduling Many-body Correlation Functions on Accelerators}
}

% \author{\IEEEauthorblockN{1\textsuperscript{st} Oguz Selvitopi}
% \and
% \and
% \and
% \and
% \and

\author{
    \IEEEauthorblockN{Oguz Selvitopi\IEEEauthorrefmark{1}, 
    Emin Ozturk\IEEEauthorrefmark{2},
    Jie Chen\IEEEauthorrefmark{4},
    Ponnuswamy Sadayappan\IEEEauthorrefmark{2},
    Robert G. Edwards\IEEEauthorrefmark{4},
    Ayd{\i}n Bulu\c{c}\IEEEauthorrefmark{1}\IEEEauthorrefmark{3}
    % David Culler\IEEEauthorrefmark{1}
    }    
    \IEEEauthorblockA{\IEEEauthorrefmark{1}Applied Mathematics and Computational Research Division, Lawrence Berkeley National Laboratory}
    \IEEEauthorblockA{\IEEEauthorrefmark{2}School of Computing, University of Utah}
    \IEEEauthorblockA{\IEEEauthorrefmark{3}Department of Electrical Engineering and Computer Sciences, University of California}
    \IEEEauthorblockA{\IEEEauthorrefmark{4}Jefferson Lab, Newport News}
}

\maketitle

\begin{abstract}
Computation of correlation functions is a key operation in Lattice quantum chromodynamics (LQCD) simulations to extract nuclear physics observables.
These functions involve many binary batch tensor contractions, each tensor possibly occupying hundreds of MBs of memory. 
Performing these contractions on GPU accelerators poses the challenge of scheduling them as to optimize tensor reuse and reduce data traffic. 
In this work we propose two fast novel scheduling algorithms that reorder contractions to increase temporal locality via input/intermediate tensor reuse.
Our schedulers take advantage of application-specific features, such as contractions being binary and locality within contraction trees, to optimize the objective of minimizing peak memory.
We integrate them into the LQCD analysis software suite Redstar and improve time-to-solution. Our schedulers attain upto 2.1x improvement in peak memory, which is reflected by a reduction of upto 4.2x in evictions, upto 1.8x in data traffic, resulting in upto 1.9x faster correlation function computation time.
% Old abstract below
% Computation of correlation functions is a key operation in Lattice quantum
% chromodynamics (QCD) simulations to extract nuclear physics observables.
% %
% These functions involve many binary batch tensor contractions.
% %
% Performing tens of thousands of contractions on GPU accelerators poses two main
% challenges: how to optimize limited device memory for the contractions and how
% to distribute the contractions among multiple accelerators as to optimize data
% transfer between devices and between host and the devices.
% %
% To address the former, we propose fast and novel scheduling algorithms
% that reorder the contractions to increase input/intermediate tensor reuse
% on a single accelerator.
% %
% We tackle the latter by proposing partitioning models that avoid
% device-to-device data transfers at the expense of increasing the number of
% tensor contractions, which is minimized by the partitioner.
% %
% Our approaches are realized within the Lattice QCD analysis software suite
% Redstar and our experiments show that the proposed methods greatly benefit the
% correlation function computations.

% Can be TBs of data moved.
% Results up to xx reduction h2d transfer
% resulting in runtime improvement

% sibling better in smaller, tree in larger
% up to 2.1x peak memory cdag

% and the decrease of the memory usage of the tensors
% could allow this type of computations to be carried out for more
% complex systems.
\end{abstract}


\section{Introduction}
\label{sec:intro}

%% PAR corr functions and LQCD
Calculations of many-particle correlation functions play an important role in producing observables in scientific physics systems such as Lattice quantum chromodynamics (LQCD) simulations~\cite{HadronSpectrum:2009krc}.
Correlation function computations for LQCD simulations involve contractions of quark propagations that describe the interactions between hadrons.
Hadronic systems consist of mesons and/or baryons, which are subatomic particles consisting of quarks held together by gluons.
These computations include the evaluation of quark propagation diagrams, whose number has factorial growth in the number of quarks and degrees of freedom of the hadronic system, making it computationally challenging to compute even simple systems such as two mesons.

%% PAR graph representation 
A natural way to represent the evaluation of hadronic correlation functions is through graphs, for which vertices are used to represent the quarks in hadrons and edges are used to represent the quark propagations.
In order to evaluate the function, a series of quark contractions is performed by eliminating one edge at a time until there are two hadrons left.
Each quark contraction necessitates a batched tensor contraction, where the dimension of the tensors ($N$) in the contraction are typically in hundreds but it is desirable to have the dimension as large as possible.
Moreover, terms of the summation used to evaluate the correlation function can be in the scale of tens of thousands.
Hence, from a computational perspective, we have thousands of graphs, each of which contain considerable amount of batched tensor contractions with individual contraction sizes reaching up to 1024, and with some of the contractions appearing in multiple graphs.

%% PAR nature of contractions
Computing tensor contractions is both compute- and memory-intensive.
To give an example, storing a single hadron node may require hundreds or thousands of MBs: a single baryon node with 64 spin components and dimension of $N=128$ will need 2 GB to store.
On an accelerator with 32 GB of memory, this will result in fewer than 16 hadron nodes that can be stored on device memory at any point of the execution.
Moreover, since the same hadron nodes appear in different graphs, data reuse becomes critical in optimizing the memory footprint.
Throughout the execution many intermediate tensors are produced, which are released as soon as there are no contractions depending on them.
Hence, it is important to perform the contractions that depend on the same tensors in a short window of time frame as to improve data reuse.
In this regard, scheduling of contractions becomes critical for efficient computation of correlation functions.

%% PAR addressed problem
In this work, we address the problem of scheduling tensor contractions needed by hadronic correlation functions on accelerators.
Although there are several works that optimize tensor contractions~\cite{Abdelfattah2016,Bibireata2004,Kim2018,Liu2021,Shi2016}, few focus on scheduling.
Due to the limited device memory, intermediate tensors may need to be moved in and out of the device memory using the PCIe link between the host and the device.
As same tensors may be needed by different contractions, an important objective of the scheduling should be maximizing tensor reuse (i.e., temporal locality) to reduce the data traffic between host and the device.
This data traffic can be a big bottleneck preventing efficient computation of correlation functions on accelerators, especially for large $N$.
When we consider the architectural trend that time to process data is reduced faster than the time to move the data, designing algorithms that focus on the latter becomes more important in performance.

%% PAR scheduling algorithms: features (application specific properties)
We propose two novel and fast algorithms for scheduling contractions that are needed by the correlation function computations in LQCD simulations.
Both of the proposed schedulers take advantage of the specific properties that stem from the application.
Our first scheduler is called the sibling scheduler and it exploits the property that each contraction is binary.
Our second scheduler is called the tree scheduler and it exploits the locality found within contraction trees to schedule them as a whole.
Sibling scheduler is faster but makes localized decisions.
On the other hand, tree scheduler is more involved and takes into account the entire state of the memory in making scheduling decisions with a well-defined gain definition.
Both schedulers make use of the contraction directed acyclic graph (DAG), which is constructed from the individual graphs that represent the correlation functions.

%% PAR Redstar integration
We realize the proposed schedulers in the LQCD analysis software suite Redstar.
Redstar solves correlation functions to obtain phsyics observables of many-particle systems.
It is capable of running on shared-memory systems and accelerators involving GPUs.
With our experiments, we demonstrate that the proposed schedulers greatly benefit the computation of correlation functions on GPUs within Redstar by optimizing data movement.
Our contributions are summarized as follows:
\begin{itemize}[leftmargin=*]
    \item We propose a scheduler, called the sibling scheduler, that exploits the property that each contraction is binary and the contraction DAG is shallow.
    \item We propose another scheduler, called the tree scheduler, that utilizes the locality inherent in contraction trees and uses the global state of the memory to make scheduling decisions.
    \item We integrate the proposed schedulers into the LQCD analysis software Redstar.
    \item We validate the proposed schedulers by assessing how well they optimize their target objective on the contraction DAGs corresponding to the hadronic correlation functions.
    \item We conduct extensive experiments within Redstar on six different correlation functions to demonstrate their effectiveness in reducing data traffic between CPU and GPU and improving time-to-solution of correlation functions.
\end{itemize}

%% PAR
The rest of the paper is organized as follows.
In Section~\ref{sec:pre}, we introduce the notation and define the problem addressed.
We describe the proposed schedulers and analyze them in Section~\ref{sec:sch}.
We present the obtained results in Section~\ref{sec:exp}.
We provide the related work in Section~\ref{sec:rw} and conclude in Section~\ref{sec:conc}.

\section{Preliminaries}
\label{sec:pre}

\subsection{Redstar software suite}
%% PAR Redstar in general
Redstar\footnote{\url{https://github.com/JeffersonLab/redstar}} is a Lattice QCD analysis software suite developed for the purpose of constructing and evaluating correlation functions on shared-memory and many-core architectures.
It manages a workflow that consists of several stages and makes use of different libraries.
A unique feature and central to the design of Redstar is the graph representation of quark propagations~\cite{Chen2023} in which hadron nodes and quarks in them are represented by vertices and quark propagations are represented by edges.
Redstar's input is a list of correlation functions that describe a physics system, from which it produces hadron nodes and associated graphs.
Using these graphs, it forms a general directed acyclic graph (DAG) that encompasses all correlation functions, and performs graph-based optimizations to efficiently generate execution queues to perform hundreds of thousands of tensor contractions.
Redstar is capable of performing contractions in parallel on CPUs and GPUs.
Among some of the libraries Redstar uses are Hadron and Colorvec, the former mainly responsible for performing tensor contractions and the latter construction of hadron nodes.

%% PAR data reuse & scheduling
\textbf{Data reuse and scheduling.}
In order to increase data reuse and reduce memory footprint, in determining a contraction path for a correlation function Redstar first pre-processes edges to compute the occurrence of each edge in all contraction trees.
Then, it assigns a value to each edge in a contraction tree using the computed frequencies and the computational complexity of the respective contraction (which can be $O(N^3), O(N^4), O(N^5)$).
Hence, in selecting a contraction path, it prefers edges that occur across different contraction trees and contractions that are expensive.
By doing so, it can avoid repeated computation of expensive contractions.
Redstar also exploits data reuse by processing contraction trees that share common contractions successively.
To do so, it sorts the contraction trees based on their similarity.
In this way, a shared tensor can be released as soon as the contraction trees that need this tensor are processed.
However, such a static and localized approach may still suffer high memory footprint because reordering of the trees does not consider the entire state of the memory, but rather similarity to another contraction tree.
In our work, we propose schedulers that aims to find contraction orders that improves data reuse and reduce peak memory.
Redstar has another scheduler~\cite{Wang2022a} for multiple GPUs but this is not focus of our work.

%% PAR Memory manager
\textbf{GPU memory management.}
Tensor contractions require large amount of memory.
Each tensor can occupy hundreds of MBs of memory and computation of the correlation functions can produce hundreds of thousands of intermediate tensors.
Limited amount of GPU memory necessitates efficient management of memory resources on the device.
For this purpose, Redstar relies on a custom memory management framework called MemHC~\cite{Wang2022b}.
MemHC includes several optimizations specifically targeting correlation functions such as duplication-aware management, lazily-released memory blocks, pre-protected LRU eviction policy, etc.
When there is not enough memory on the device for a contraction, MemHC evicts memory blocks to host memory.
If the scheduling order of contractions has poor tensor reuse, this will result in a lot of data movement between the host and the device due to evictions.
The data moved from host to device can go up to a few TBs, making the PCI bus a bottleneck.
For example, the default sorting-based contraction scheduler in Redstar causes more than 7 TB of data traffic between host and device on the \texttt{roper} dataset for the tensor dimension of 64.
The proposed schedulers in this work aim to overcome this bottleneck.

%% PAR contractions in Redstar
% Redstar has various interfaces to support operations related to the contractions.
% %
% These operations can mainly be categorized into three as: (i) hadron exterior contractions (HEC), which are used in the generation of intermediate tensors, (ii) hadron contract all operations (HCA), which are used in the generation of the final values, and (iii) deletion of tensors, which are utilized when an intermediate tensor is not needed and consists of freeing the memory used by the tensor.
% %
% The number of HEC and HCA type of operations may vary according to the correlation function being computed.
% %
% HEC-type of operations are more compute-intensive than the HCA-type of operations with the latter being more memory-bound.

\subsection{Notation}
%% PAR: Contraction trees 
The input to the computations of correlation functions is a set of $k$ contraction trees, denoted with $\T = \{T_1, T_2, \cdots, T_k\}$, which are rooted and directed.
The node sets of a pair of trees in $\T$ need not be disjoint except the roots of these trees.
Each tree constructed for the computation of correlation functions in our application typically has a few nodes in it, usually no more than 10 or 15, and there are many contraction trees whose number can range from a few thousand to hundreds of thousands, depending on the type and parameters of LQCD simulations.

\begin{figure}[t]
  \begin{center}
\includegraphics[width=0.30\textwidth]{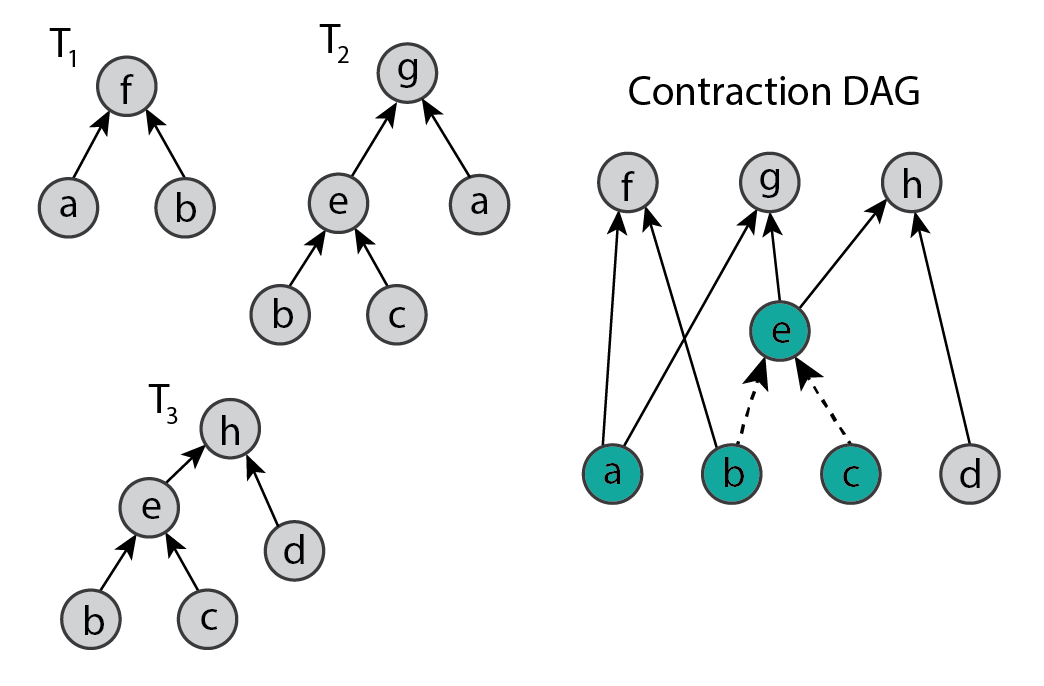}
\vspace{-2ex}
\caption{The contraction DAG formed from three contraction trees. Nodes that appear in multiple trees are shown in green color. The edges that appear in multiple trees are shown with dashed arrows.}
\label{fig:cdag-form}
\end{center}
\vspace{-4ex}
\end{figure}

%% PAR: Contraction DAG and node
From $\T$, a contraction DAG $G = (V, E)$ is formed, in which each node $u \in V$ represents a tensor contraction and/or a tensor, and each directed edge $(u, v) \in E$ represents the dependency of the contraction represented by $v$ to the tensor represented by $u$.
The leaf nodes in $V$ represent only a tensor, while the non-leaf nodes represent both a tensor contraction and the output tensor produced by that contraction.
Fig.~\ref{fig:cdag-form} shows an example contraction DAG formed from three contraction trees.
%
% \todo{Each tensor contraction is binary, hence, each non-leaf node has exactly two children.}
%
% don't mention mts a new node type
We have the following fields for each node $u$:
\begin{itemize}[leftmargin=*]
    % \item $v_i.left$: First input tensor of a contraction (for non-leaf nodes).    
    % \item $v_i.right$: Second input tensor of a contraction (for non-leaf nodes).
    \item $u.child$: Inputs of a tensor contraction (for non-leaf nodes). 
    We also use $u.left$ and $u.right$ to refer to the specific child of a node as in context as there are only two children of a non-leaf node $u$.
    
    \item $u.parents$: The contractions that depend on this tensor. For the roots of trees in $\T$, $u.parents = \emptyset$.
    \item $u.type$: The type of a node can be one of the following: \texttt{LEAF}, \texttt{INTERIOR}, \texttt{ROOT}. 
    The nodes that do not have any input edges are of type \texttt{LEAF}, whereas the nodes that do not have any output edges are of type \texttt{ROOT} and there is one root for every contraction tree $T_i$, i.e., there are $k$ \texttt{ROOT} nodes.
    The remaining set of nodes are \texttt{INTERIOR} nodes that correspond to the contractions which need to be scheduled and their outputs need to be stored as long as they are needed.
    The total number of contractions hence is given by the number of \texttt{INTERIOR} and \texttt{ROOT} nodes, i.e., number of non-leaf nodes.

    \item $u.cost$: The computational cost of the binary contraction.
    
    \item $u.size$: The size of the tensor represented by $u$ in memory.
\end{itemize}
% We note that there is a small set of contractions that appear in multiple time-slices.
%
% These contractions are computed at the beginning and made available throughout the execution.
%
% Such nodes are not part of the scheduling and they are treated similar to leaf nodes.
%
Finally, there is a weight associated with each edge $(u, v)$, denoted with $w(u,v)$, which is equal to the size of the tensor needed by the contraction represented by $v$, i.e., $w(u, v) = u.size$.

Finally, the proposed algorithms are easily generalized to the general case where each node has arbitrary number children without affecting their runtime complexity.
We use the notation $u.left$ and $u.right$ when the scheduler exploits the property that each node has two children (i.e., sibling scheduler in Section~\ref{sec:sibling}), and the notation $u.child$ otherwise (i.e, tree scheduler in Section~\ref{sec:tree}).

% 2~3 paragraphs
% \todo{+figure explaining a contraction}
% hadron contractions
% meson baryons

\subsection{Problem definition}
\label{sec:pre-problem}
%% PAR memory model
We first describe the memory model we use for scheduling the nodes of the contraction DAG.
A tensor remains in memory as long as it is needed by a contraction that is not yet performed.
At the beginning, we assume the tensors that correspond to the leaf nodes reside in the host memory and not in device memory.
Hence, a tensor corresponding to a leaf node is not considered to be consuming memory until the first contraction that depends on that leaf is performed.
Consider a sequence of $n$ contractions $c_1, c_2, \cdots, c_n$ and let $M_{i-1}$ be the amount of memory used before performing $c_i$, where $M_0 = 0$.
Processing $c_i$ has the following steps: (i) bring any tensor that correspond to leaf nodes that $c_i$ depends on to memory, if any, (ii) perform $c_i$ and produce its output tensor, and (iii) release the tensors that do not have any contractions depending on them, including the one produced by $c_i$.
Note that the tensors that correspond to interior nodes that $c_i$ depend on are already in memory before performing $c_i$.
After these steps, the used memory is $M_i$.
The peak memory in this sequence of contractions is given by $\max_i{M_i}$ and the memory change due to performing $c_i$ is $M_i-M_{i-1}$.
Note that $M_n=0$.
Table~\ref{tb:sch-mem} presents the contents of the memory for two different contraction schedules $S_1$ and $S_2$ on the contraction DAG in Fig.~\ref{fig:cdag-form}.

\begin{table}[t]
  \caption{Memory used in two different schedules on the contraction DAG in Fig.~\ref{fig:cdag-form}.}
  \vspace{-3ex}
  \begin{center}
    \scalebox{0.80} {
      \begin{tabular}{c c c c r}
        \toprule
         & & \multicolumn{2}{c}{Memory contents} &   \\
          \cmidrule(lr){3-4} 
        Schedule & Contraction & Before & After & Size \\
        \midrule
	\multirow{4}{*}{$S_1$} & $e$ & $\emptyset$ & $\{b,e\}$ & 2 \\
                           & $g$ &  $\{b,e\}$ & $\{a,b,e\}$ & 3 \\
                           & $h$ & $\{a,b,e\}$ & $\{a,b\}$ & 2 \\
                           & $f$ & $\{a,b\}$ & $\emptyset$ & 0 \\
        \midrule
        \midrule
        \multirow{4}{*}{$S_2$} & $f$ & $\emptyset$ & $\{a,b\}$ & 2 \\
                           & $e$ &  $\{a,b\}$ & $\{a,e\}$ & 2 \\
                           & $g$ & $\{a,e\}$ & $\{a\}$ & 1 \\
                           & $h$ & $\{a\}$ & $\emptyset$ & 0 \\
	  \bottomrule
        \end{tabular}
        } 
  \end{center}
  \label{tb:sch-mem}  
  \vspace{-3ex}
\end{table} 

%% PAR goal
We address the problem of finding a sequential schedule of the contraction DAG that minimizes the peak memory, i.e., that minimizes $\max_i{M_i}$.
A good scheduler should intuitively maximize data reuse in order to reduce peak memory.
Reducing the peak memory should reduce the evictions in Redstar described earlier, resulting in reduced data transfers between host and device, and hence improved runtime for the computation of correlation functions.
\section{Scheduling heuristics}
\label{sec:sch}
In this section propose two schedulers, sibling scheduler and tree scheduler, to solve the problem defined in Section~\ref{sec:pre-problem}.

\subsection{Sibling scheduler}
\label{sec:sibling}

%% PAR Motivation
The sibling scheduler aims to exploit the structural property of the contraction DAG that each contraction is binary.
As soon as the contraction represented by $u$ is complete and its output is produced, it strives to complete the contractions represented by each node $v$ that are sibling of $u$ as to enable the parent contractions.
The scheduler gives a higher priority to the available contractions that are further in the topological ordering of $G$ and in this manner aims to achieve a depth-first scheduling of the contractions to reduce the memory footprint of the stored intermediate tensors.

%% PAR Structures and mid-level details
The sibling-based scheduler computes and/or maintains the following information associated with each vertex $u$:
\begin{itemize}[leftmargin=*]
    \item $u.rank$: Rank of each node in the contraction DAG, where
    \begin{equation}
        u.rank =
        \begin{cases}
        0 & \text{if $u$ is \texttt{LEAF},}\\
        \max\limits_{v \in u.child}{(v.rank)} + 1 & \text{otherwise.}
        \end{cases}       
    \end{equation}
    Ranks are computed prior to scheduling and does not change throughout the scheduling of the contractions.

    \item $u.rs$: Number of contractions that depend on the tensor represented by $u$ and that are not yet performed, i.e., number of remaining successors to enable the release of $u$.
    Initially, $u.rs = |u.parents|$.

    \item $u.rp$: Number of tensors that the contraction represented by $u$ depends on and that are not yet available, i.e., number of remaining predecessors to enable $u$ for contracting.
    Initially, $u.rp = |u.child|$.

    \item $u.state$: There are four states that may be associated with nodes: \texttt{WAITING}, \texttt{QUEUED}, \texttt{INMEM}, and \texttt{RELEASED}.
    Initially, all nodes are in state \texttt{WAITING}.
\end{itemize}

%% PAR fields rs and rp
The field $u.rs$ is maintained to check whether the tensor represented by $u$ is releasable from memory upon completing a contraction that depends on it, i.e., it becomes releasable when $u.rs = 0$ where there is not any contraction task that depends on this tensor.
On the other hand, the field $u.rp$ is maintained to see whether we can enqueue the respective contraction operation upon a child of $u$ being contracted and hence an input tensor for $u$ being available.

\setlength{\textfloatsep}{0pt}
\begin{algorithm}[t]
\caption{SB-SCHEDULER ($G=(V,E)$)}
\label{alg:sibling-schedule}
\begin{algorithmic}[1]
\footnotesize
\STATE Let $Q_1, Q_2, \cdots, Q_q$ be empty queues
\WHILE{There are contractions to schedule}
  \IF{All queues are empty}
    \STATE Let $u$ be a random leaf with $u.state = $ \texttt{WAITING}
  \ELSE
    \STATE Let $i$ be the largest integer such that $Q_i \neq \emptyset$
    \STATE $u$ = DEQUEUE($Q_i$)
  \ENDIF
  \STATE SB-PROCESS($u$)
\ENDWHILE

\end{algorithmic}
\end{algorithm}

%% PAR field state
The state field has different semantics depending on whether node is leaf or not.
A leaf node is in state (i) \texttt{WAITING} if the respective tensor is not brought to memory yet, (ii) \texttt{INMEM} if the respective tensor is in memory, and (iii) \texttt{RELEASED} if the respective tensor is not needed any more and released from memory.
A non-leaf node, on the other hand, is in state (i) \texttt{WAITING} if the respective contraction is not yet available for computing, (ii) \texttt{QUEUED} if the respective contraction can be scheduled and is in one of the queues, (iii) \texttt{INMEM} if the output tensor generated by the respective contraction resides in memory, and (iv) \texttt{RELEASED} if the respective tensor is not needed any more and released from memory.
The state \texttt{QUEUED} cannot be used by leaf nodes as it is for the nodes that signify a contraction, i.e., non-leaf nodes.

%% PAR queues
The sibling-based scheduler maintains a number of queues throughout the scheduling process to distinguish between contractions of different ranks.
Excluding the leaf nodes, we maintain $q = max_{u \in V}{u.rank}$ queues $Q_1, Q_2, \cdots, Q_q$, where $Q_i$ contains the nodes $v.rank = i$.
The higher the rank of a contraction, the higher its priority, i.e., the contractions in $Q_i$ has a higher priority than the contractions in $Q_j$, where $i > j$.

%% PAR Algorithms and descriptions: schedule
The sibling scheduler given in Alg.~\ref{alg:sibling-schedule} utilizes two helper routines SB-PROCESS and SB-PROP-DOWN, presented in Alg.~\ref{alg:sibling-process} and Alg.~\ref{alg:sibling-propdown}, respectively.
In each iteration of the scheduler in Alg.~\ref{alg:sibling-schedule}, we select a contraction task from the highest level queue available.
If all the queues are empty yet there are still contraction tasks that need to be scheduled (which may happen at the beginning of the scheduling process or due to the structure of the contraction DAG), we select a random leaf node.
Then we perform some operations triggered by the selection of that node by calling SB-PROCESS.

\begin{algorithm}[t]
\caption{SB-PROCESS ($u$)}
\label{alg:sibling-process}
\begin{algorithmic}[1]
\footnotesize

\IF {$u.type \neq$ \texttt{LEAF}}
  \STATE Perform the contraction corresponding to $u$
\ELSE
  \STATE Bring the tensor represented by $u$ to memory \vskip 4pt
\ENDIF

\STATE $u.state =$ \texttt{INMEM} \vskip 4pt

\STATEx // Check for releasable nodes
\IF {$u.type \neq$ \texttt{LEAF}}
  \FOR {$v \in u.child$}
    \STATE $v.rs = v.rs - 1$
    \IF {$v.rs == 0$}
      \STATE $v.state = $ \texttt{RELEASED}
    \ENDIF
  \ENDFOR
  
  \IF {$u.type == $ \texttt{ROOT}}
    \STATE $u.state =$ \texttt{RELEASED} \vskip 4pt
  \ENDIF
\ENDIF

\STATEx // Process siblings or enqueue parents
\FOR{$v \in u.parents$}
  \STATE $v.rp = v.rp - 1$
  \IF {$v.rp == 1$}
    \STATE Let $w$ with parent $v$ be the sibling of $u$
    \IF{$w.state == $ \texttt{WAITING}}
      \STATE SB-PROP-DOWN($w$)
    \ENDIF
  \ELSIF{$v.rp == 0$}
    \STATE ENQUEUE($Q_{v.rank}$, $v$)
    \STATE $v.state = $ \texttt{QUEUED}
  \ENDIF
\ENDFOR

\end{algorithmic}
\end{algorithm}

%% PAR Alg PROCESS-NODE 
The SB-PROCESS in Alg.~\ref{alg:sibling-process} performs the operations after a tensor becomes available in memory, which can happen either by bringing a tensor that corresponds to a leaf in the contraction DAG (line 4 of Alg.~\ref{alg:sibling-process}) or completing a contraction task represented by a non-leaf node (line 2 of Alg.~\ref{alg:sibling-process}).
These operations can be classified into two.
In the first (lines 6-12 of Alg.~\ref{alg:sibling-process}), we first check for any releasable tensors from memory after completing a contraction task $u$ corresponding to a non-leaf node by traversing $u$'s children, and updating and checking their $rs$ field.
Note that the nodes with \texttt{ROOT} type are immediately releasable after being contracted as there are no contraction tasks that depend on them (line 11-12 of Alg.~\ref{alg:sibling-process}).
The second kind of operations involve processing siblings or enqueueing parent nodes of $u$ (lines 13-21 of Alg.~\ref{alg:sibling-process}).
For parent $v$ of $u$, if $u$'s respective sibling $w$ is not yet contracted and hence the corresponding output tensor is not available, we try to make this sibling available by propagating down all the nodes in the sub-tree whose root is $w$ (line 18 of Alg.~\ref{alg:sibling-process}) by calling SB-PROP-DOWN.
Otherwise, both tensors represented by $u$ and $w$ are available and in memory, hence the contraction represented by parent $v$ can now be performed.
Thus, we enqueue the corresponding contraction task to the respective queue (line 20 of Alg.~\ref{alg:sibling-process}).
The operations ENQUEUE and DEQUEUE on the queues are self-explanatory.
% \todo{+Figure explaining the operations after contracting a node}

%% PAR Alg PROP-DOWN
The SB-PROP-DOWN in Alg.~\ref{alg:sibling-propdown} aims to make the sibling of a node available as to enable their parent contraction as soon as possible.
For this purpose, the leaf descendants of the sibling are brought to memory if they are not already in memory.
The call to SB-PROP-DOWN is done only if there is one remaining predecessor of the parent node (line 15 of Alg.~\ref{alg:sibling-process}).
This is expected to be more effective especially when each contraction is binary.
% \todo{PROP-DOWN still specific to two-child DAG.}

%% PAR Analysis/runtime
\emph{\textbf{Runtime complexity.}}
In the sibling scheduler, the routine SB-PROCESS can be called on leaf nodes whose state is \texttt{WAITING} and for the non-leaf nodes whose state is \texttt{QUEUED}.
Once SB-PROCESS is called for node $u$, its state permanently changes to either \texttt{INMEM} or \texttt{RELEASED}.
Hence SB-PROCESS is called only once for each node.
Once called for $u$, this routine traverses incoming and outgoing edges of $u$.
Hence, runtime complexity of sibling scheduler is $O(V+E)$.

%% Sibling-based scheduler PROP-DOWN
\begin{algorithm}[t]
\caption{SB-PROP-DOWN ($u$)}
\label{alg:sibling-propdown}
\begin{algorithmic}[1]
\footnotesize

\IF {$u.state \neq$ \texttt{WAITING}} 
  \State \textbf{return}
\ENDIF

\IF {$u.type ==$ \texttt{LEAF}}
  \STATE SB-PROCESS($u$) 
  \State \textbf{return} \vskip 4pt
\ENDIF

\STATE SB-PROP-DOWN($u.left$)
\STATE SB-PROP-DOWN($u.right$)
\end{algorithmic}
\end{algorithm}

\subsection{Tree scheduler}
\label{sec:tree}

%% PAR motivation 
Contraction trees harbor a certain degree of locality within themselves as all the tensors produced within a tree will be needed by the contractions in that tree.
Hence, we can take advantage of this specific feature provided by the input trees to schedule contractions on a coarse-grain level by scheduling contraction trees as a whole.
%
% Recall that since same tensors and/or contractions may appear in different trees, after completing the contractions in a sequence of trees $T_1, \cdots, T_i$, selection of the next tree for contracting will impact how much more memory will be used depending on the tensors the selected tree can reuse.
%
To this end, we propose a scheduler based on selecting contraction trees as a whole that utilizes the global information about the most recent state of the memory.
A contraction operation may require its input tensors brought to memory which can stay in memory after the contraction if there are other contractions depending on them, increasing the utilized memory; or it may cause its input tensors to be released from memory if they are already in memory and do not depend on any other contraction, decreasing the utilized memory.
The proposed tree scheduler efficiently maintains how much increase the subset of contractions in a tree will cause in memory and selects the next tree based on this information.
Note that the memory increase can be negative, which means a reduction in memory consumption.

\begin{figure}[t]
  \begin{center}
\includegraphics[width=0.25\textwidth]{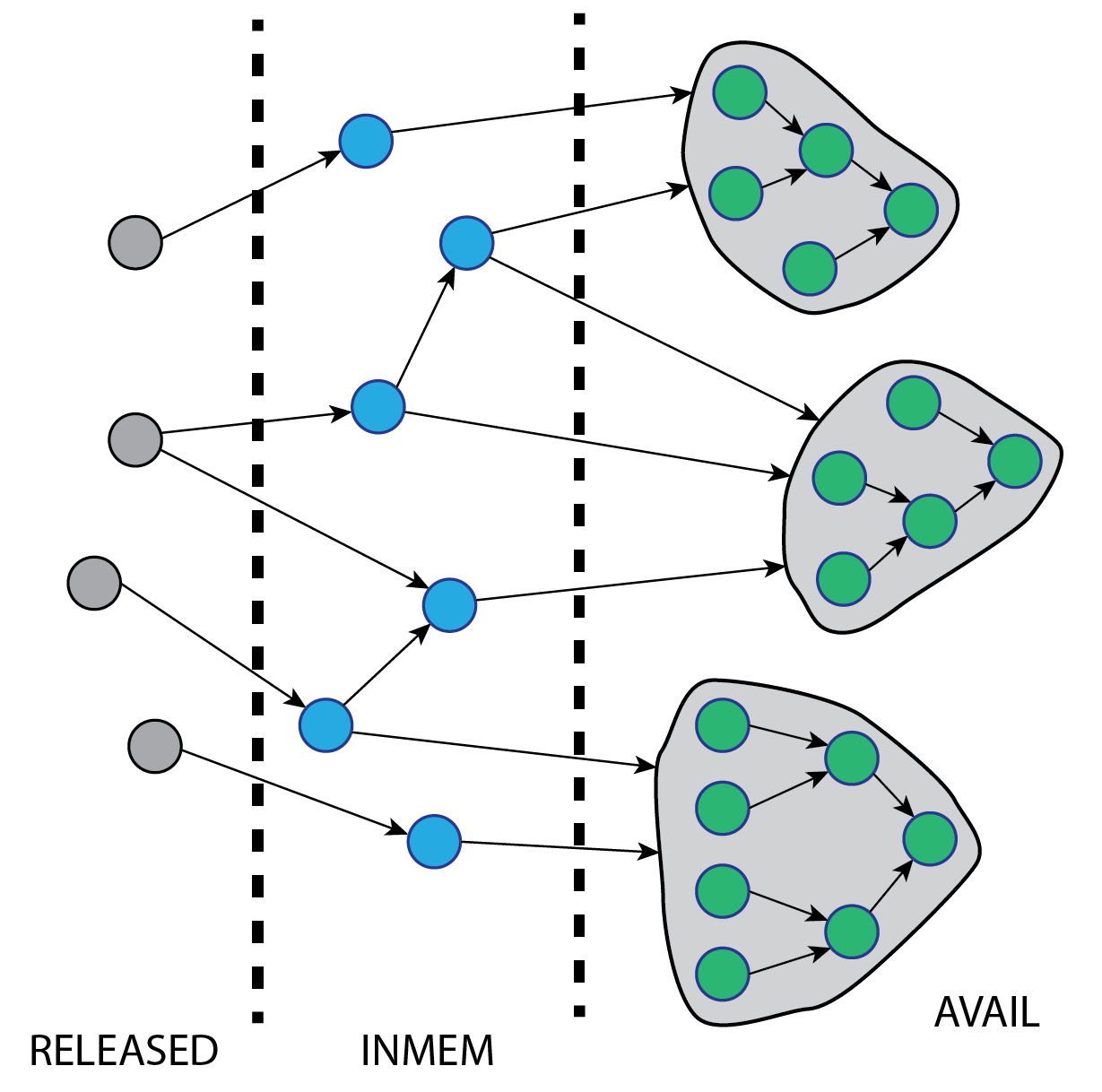}
\vspace{-1ex}
\caption{At each iteration, tree scheduler selects a contraction tree (shown on the right), then processes the nodes in that contraction tree, and then either keeps tensors in memory if they are needed by other contraction trees (shown in the middle) or releases them from memory if they are not needed by any other contractions (shown on the left).}
\label{fig:tb-overview}
\end{center}
\end{figure}

%% PAR remarks
The tree scheduler does not rely upon each non-leaf node having two children and can handle contraction DAGs in which nodes have arbitrary number of children.
Moreover, although we describe our algorithms for contraction trees, our scheduler also works where each $T_i$ is not a tree but a DAG.
Fig.~\ref{fig:tb-overview} illustrates the high-level concepts of the tree scheduler.

%% PAR Fields and functions for vertices
The tree scheduler maintains information on both node and tree level.
For each vertex $u$, we have the following:
\begin{itemize}[leftmargin=*]
    \item $u.ctree$: The set of trees node $u$ is in.
    
    \item $u.state$: There are three states that may be associated with nodes: \texttt{AVAIL}, \texttt{INMEM}, and \texttt{RELEASED}.
    All nodes are initially in state \texttt{AVAIL}.

    \item $u.outAv$: The set of outgoing neighbors of $u$ that is in state \texttt{AVAIL}, i.e., $\{v: v \in u.parents \mbox{ and } v.state = \texttt{AVAIL} \}$.

    \item $\tau(u, T_i)$: Number of outgoing neighbors of $u$ that are in state \texttt{AVAIL} and \emph{in} $T_i$, i.e., $| \{v: v \in u.parents \mbox{ and } v.state = \texttt{AVAIL} \mbox{ and } v \in T_i \} |$.

    \item $\delta(u, T_i)$: Number of outgoing neighbors of $u$ that are in state \texttt{AVAIL} and \emph{not in} $T_i$, i.e., $| \{v: v \in u.parents \mbox{ and } v.state = \texttt{AVAIL} \mbox{ and } v \notin T_i \} |$.
\end{itemize}

%% PAR Explain vertex fields
A node is in state (i) \texttt{AVAIL} if the respective tensor has not been brought to memory yet or the respective contraction has not been performed yet, (ii) \texttt{INMEM} if the respective tensor is in memory, and (iii) \texttt{RELEASED} if the respective tensor is not needed any more and released from memory. 
We maintain $u.ctree$ as $u$ may appear in more than one contraction tree.
Observe that $\tau(u, T_i) + \delta(u, T_i) = |u.outAv|$.
The purpose of these will be clear later, but in brief the functions $\tau$ and $\delta$ allow us to perform necessary updates efficiently.
Among these fields, $u.ctree$ is used for nodes in state \texttt{AVAIL} and the rest of them are used for nodes in state \texttt{INMEM}.

%% PAR tree fields
We introduce the concept of \emph{gain}, which is used to signify the change in memory if a contraction or subset of contractions were to be performed.
Gain is positive if memory consumption decreases and negative if memory consumption increases.
If the utilized memory space before contracting $T_i$ is $M$, and it becomes $M^{\prime}$ after contracting $T_i$, the gain of $T_i$ is given by $M - M^{\prime}$.
We do not include temporary memory that might be needed while processing the contractions in $T_i$.
We associate gains with nodes as well as trees.
However, the choices are made on the gain values of trees.
For each tree $T_i$, we maintain the following:
\begin{itemize}[leftmargin=*]
    \item $T_i.pred$: The set of vertices that have an outgoing neighbor in $T_i$ and in state \texttt{INMEM}, i.e., $\{u : \exists v \in T_i \mbox{ and } (u,v) \in E \mbox{ and } u.state = \texttt{INMEM} \}$.

    \item $T_i.u.igain$: The individual gain signifying the memory increase or decrease if the contraction represented by $u \in T_i$ were performed.

    \item $T_i.cgain$: The coarse gain (i.e., $T_i$ as a whole) due to the amount of tensors released from memory if  contractions in $T_i$ were performed.

    \item $T_i.tgain$: The overall gain signifying the memory increase or decrease if contractions in $T_i$ were performed.
    This is equal to the summation of individual gains and coarse gain, i.e., $T_i.cgain + \sum_{u \in T_i} T_i.u.igain$.
\end{itemize}

%% PAR remarks about tree fields
$T_i.u$ indicates node $u$ in $T_i$.
We maintain the predecessors of $T_i$, $T_i.pred$, in order to determine whether performing the contractions in $T_i$ enables release of any tensor corresponding to a node in $T_i.pred$.
Among the gains associated with each tree, only the overall gain, i.e., $T_i.tgain$, is used in choosing the trees for contracting.
The rest of the gain fields are kept in order to make the description of the algorithms easier.

\begin{algorithm}[t]
\caption{TR-SCHEDULER ($G=(V,E)$, $\T = \{T_1, T_2, \cdots, T_k\}$)}
\label{alg:tree-sch}
\begin{algorithmic}[1]
\footnotesize

\STATE TR-INIT($G, \T$)
\STATE $A = \T$
\WHILE{$A \neq \emptyset$}
  \STATE $T^{\prime} = \max_{T_i \in A} T_i.tgain$
  \STATE PROCESS-CTREE($T^{\prime}$)
  \STATE $A = A - T^{\prime}$
\ENDWHILE

\end{algorithmic}
\end{algorithm}

%% PAR How the scheduler works overall
% Brute-force vs. efficiently updating fields
The tree scheduler selects the contraction tree $T_i$ with the highest gain (using the field $T_i.tgain$) at each iteration among the trees that are not yet contracted (Alg.~\ref{alg:tree-sch}).
In that sense, we determine a scheduling order for the contraction trees.
After processing the nodes in $T_i$, some of the tensors produced may remain in memory and some of them may be released depending on whether there are contractions in other trees that depend on them.
The processing of $T_i$ may necessitate gain updates of other trees that are not yet contracted.
This can be done by computing all gains from scratch.
However, this is prohibitively expensive.
We show that the proposed algorithms can perform the updates much faster with the help of the fields maintained and described above.

%% PAR Main sketch of the algorithm
% Which portions are critical
Tree scheduler first initializes the described fields with the routine TR-INIT (Alg.~\ref{alg:tr-init}) and then repeatedly selects a tree whose contraction leads to minimum increase (or maximum decrease) in memory.
Processing of a contraction tree may lead to gain updates of other contraction trees.
This is handled with the routine PROCESS-CTREE (Alg.~\ref{alg:process-ctree}) along with necessary field changes due to the operations that result from performing contractions in the selected tree.

%% INIT
\begin{algorithm}[t]
\caption{TR-INIT ($G=(V,E)$, $\T = \{T_1, T_2, \cdots, T_k\}$)}
\label{alg:tr-init}
\begin{algorithmic}[1]
\footnotesize

\FOR{$u \in V$}
  \STATE $u.outAv = u.parents$  
  \STATE $u.ctree = \emptyset$
  \STATE $u.state = \texttt{AVAIL}$ \vskip 4pt
\ENDFOR

\FOR{$T_i \in \T$}
  \STATE $T_i.pred = \emptyset$  
  \STATE $T_i.cgain = 0$,  $T_i.tgain = 0$
  \FOR{$u \in T_i$}
    \STATE $u.ctree = u.ctree \cup T_i$ \vskip 4pt
    % \STATE $\tau(u, T_i) = 0$, $\delta(u, T_i) = 0$ 
  \ENDFOR
\ENDFOR

\STATEx // Initialize individual gains
\STATE $S = \emptyset$
\FOR{$T_i \in \T$}
  \FOR{$u \in T_i$}
    \STATE $g(u, T_i) = |outAv(u)|$
    \STATE $S = S \cup g(u, T_i)$ \vskip 4pt
  \ENDFOR
\ENDFOR

\FOR{$(u,v) \in E$}
  \FOR{$T_i \in u.ctree$}
    \IF{$T_i \in v.ctree$}
      \STATE $g(u, T_i) = g(u, T_i) - 1$ \vskip 4pt
    \ENDIF
  \ENDFOR
\ENDFOR

\FOR{$g(u,T_i) \in S$}
  \IF{$g(u,T_i) == 0$}
    \STATE $T_i.u.igain = 0$
  \ELSE
    \STATE $T_i.u.igain = -u.size$
  \ENDIF
  \STATE $T_i.tgain = T_i.tgain + g(u,T_i)$
\ENDFOR
 
\end{algorithmic}
\end{algorithm}

%% PAR Init
\emph{\textbf{Initialization.}}
In Alg.~\ref{alg:tr-init} we initialize the fields for nodes and contraction trees besides the initial contraction tree gains.
%
% Initialization of various fields between lines 1-11 is straightforward.
%
For the initialization of gains for $T_i$, we only have individual node gains because contracting $T_i$ will not cause any tensor that is in memory to be released as the memory is empty at the beginning.
Hence, $T_i.cgain = 0$.
Recall that the individual gain of a node $u \in T_i$ determines whether the tensor represented by $u$ will remain in memory after processing $T_i$.
The tensor represented by $u$ will be kept in memory if there are contractions in other trees that depend on this tensor, in which case this individual gain is equal to $-u.size$; otherwise it is 0.
The individual gains are computed between lines 10-24 of Alg.~\ref{alg:tr-init} and the initial gain $T_i.tgain$ of $T_i$ is set to the summation of these individual gains.

\begin{figure}[b]
  \begin{center}
\includegraphics[width=0.23\textwidth]{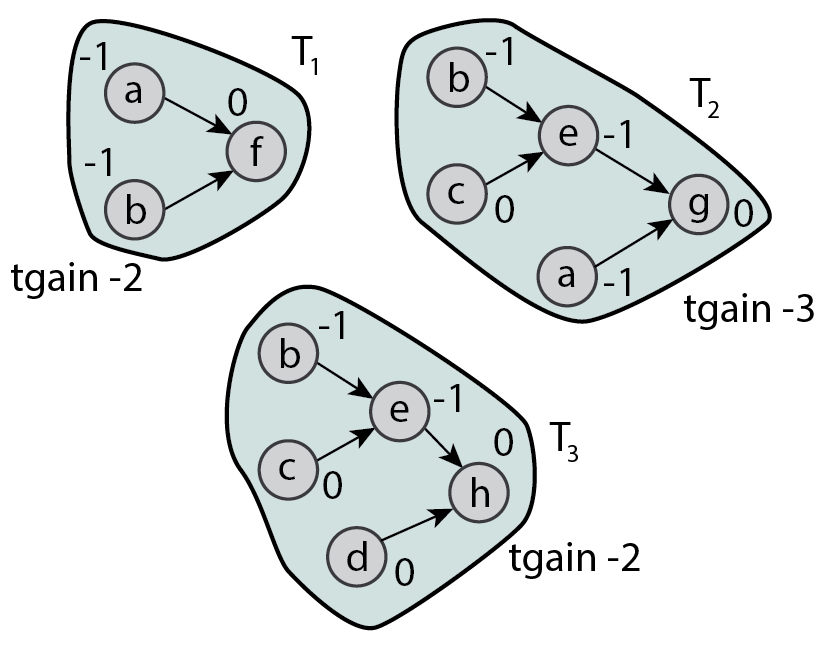}
\caption{Individual gains of nodes in trees after initialization, indicated with the numbers near the nodes. The example assumes the nodes have unit sizes.}
\label{fig:tb-igains}
\end{center}
\end{figure}

%% PAR Init example
Fig.~\ref{fig:tb-igains} illustrates the individual gains of various nodes in three trees after initialization.
%
% Observe that the same nodes or edges may appear in different trees.
% %
% Assuming all nodes have unit sizes, node $b$'s $igain$ is -1 in all trees since in whichever tree this node is processed first, the corresponding tensor needs to remain in the memory as there will be another contraction depending on it (either $e$ or $f$).
% %
% A node may appear in multiple trees yet may be releasable upon being processed.
% %
% Consider node $c$ in $T_2$.
% %
% After processing all the nodes in $T_2$, the tensor represented by $c$ will not be needed by any other contractions in other trees, as the contraction represented by node $e$ will also be produced by $T_2$.
%
Initial $T_i.tgain$ values are equal to the summation of individual gains of these nodes.

%% PAR process-ctree
\emph{\textbf{Processing a selected contraction tree.}}
After selecting a contraction tree $T$, Alg.~\ref{alg:process-ctree} processes the nodes in $T$ in topologically sorted order from left to right.
It makes use of two helper routines for processing a node $u \in T$, PROCESS-CHILD and PROCESS-NODE, the former for updates related to child $v$ of $u$ and the latter for $u$ itself.
Processing a node $u$ in Alg.~\ref{alg:process-ctree} changes $u$'s state from \texttt{AVAIL} to either \texttt{INMEM} or \texttt{RELEASED} after making necessary fields updates and possible gain updates due to its children.
The topological order utilized to process nodes in $T$ ensures that upon calling PROCESS-CHILD for child $v$ of $u$, the state of $v$ is \texttt{INMEM} while the state of $u$ is \texttt{AVAIL}.
%
% The updates regarding PROCESS-CHILD is more involved than the updates regarding PROCESS-NODE as there are several cases that can trigger a gain update for child $v$ of $u$ as $u$ is being processed and its state changing.

%% PROCESS-CTREE
\begin{algorithm}[t]
\caption{PROCESS-CTREE ($T$)}
\label{alg:process-ctree}
\begin{algorithmic}[1]
\footnotesize

\FOR{$u \in T$ in topologically sorted order}
  \IF{$u.type \neq$ \texttt{LEAF}}
    \FOR{$v \in u.child$}
      \STATE PROCESS-CHILD($u, v$)
    \ENDFOR
  \ENDIF
  \STATE PROCESS-NODE($u$)
\ENDFOR
    
\end{algorithmic}
\end{algorithm}

%% PROCESS-CHILD
\begin{algorithm}[t]
\caption{PROCESS-CHILD ($u, x$)}
\label{alg:process-child}
\begin{algorithmic}[1]
\footnotesize

\FOR{$T_i: x \in T_i.pred$}
  \IF{$u \in T_i$}    
    \IF{$\tau(x, T_i) == 1$ and $\delta(x, T_i) == 0$}
      %\Statex \hspace{\algorithmicindent} \hspace{\algorithmicindent}  // Contracting $T_i$ will not cause release of $x$ any more
      \State $T_i.cgain = T_i.cgain - x.size$
      \STATE $T_i.tgain = T_i.tgain - x.size$
    \ENDIF
    \STATE $\tau(x, T_i) = \tau(x, T_i) - 1$
    \IF{$\tau(x,T_i) = 0$}
      \STATE $T_i.pred = T_i.pred - x$
    \ENDIF
  \ELSE    
    \IF{$\delta(x,T_i) == 1$}    
      %\Statex \hspace{\algorithmicindent} \hspace{\algorithmicindent}  // Contracting $T_i$ will now cause release of $x$
      \STATE $T_i.cgain = T_i.cgain + x.size$
      \STATE $T_i.tgain = T_i.tgain + x.size$
    \ENDIF
    \STATE $\delta(x,T_i) = \delta(x,T_i) - 1$  \vskip 4pt
  \ENDIF
\ENDFOR

\STATE $x.outAv = v.outAv - u$
\IF{$x.outAv == \emptyset$}
  %\STATEx \hspace{\algorithmicindent} // Can release $x$ as nothing depends on it
  \STATE $x.state =$ \texttt{RELEASED}
\ENDIF
    
\end{algorithmic}
\end{algorithm}

\begin{figure}[b]
  \begin{center}
\includegraphics[width=0.40\textwidth]{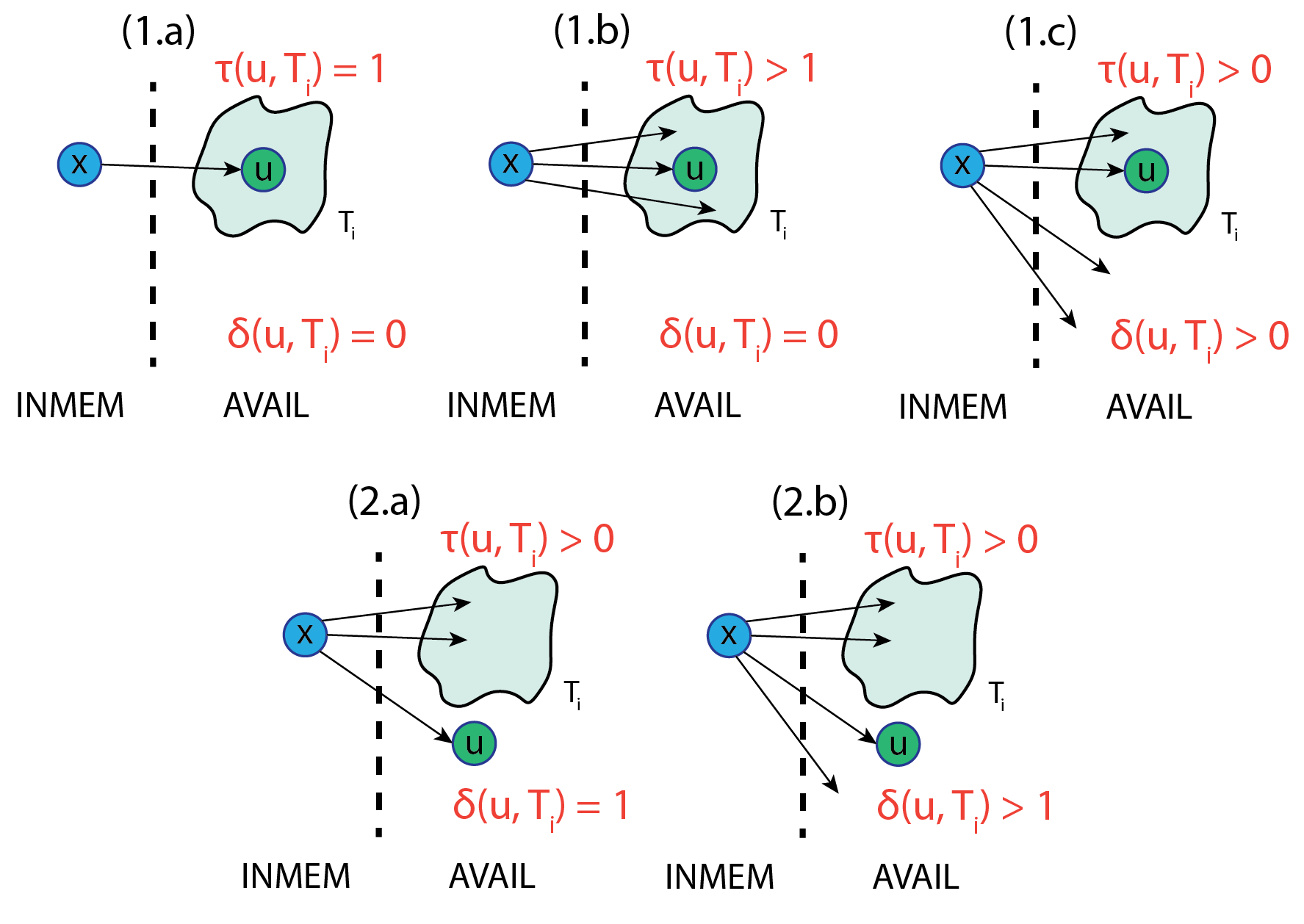}
\caption{Various cases that may trigger a gain update regarding node $x$ whose parent $u$ is being processed.}
\label{fig:tb-cases}
\end{center}
\vspace{-3ex}
\end{figure}

%% PAR process child 2~3 paragraphs
\emph{\textbf{Updates due to child $x$ of a node $u$ being processed in the selected contraction tree.}}
Contracting a node $u \in T_i$ may trigger gain updates due to its child $x$ as $u$'s state changes from \texttt{AVAIL} to \texttt{INMEM}.
Observe that $x.state =$ \texttt{INMEM}, otherwise $u$ could not have been processed due to the topological ordering used in Alg.~\ref{alg:process-ctree}.
In order to handle such cases, we present Alg.~\ref{alg:process-child}, which makes the necessary gain updates for each successor tree of $x$, i.e., the set of trees which has $x$ as their predecessor.
To perform these updates efficiently, we make use of the functions $\tau$ and $\delta$.
There are a total of five cases of interest, two of which trigger a gain update.
We consider each $T_i$ where $x \in T_i.pred$.
Note that $u$ may or may not be in $T_i$ and that $\tau(x, T_i) > 0$ as $x$ is a predecessor of $T_i$.
We categorize these five cases into two, where in three of them $u \in T_i$ and in two of them $u \notin T_i$.
These cases are also illustrated in Fig.~\ref{fig:tb-cases}.
Among these five cases, the cases (1.b), (1.c), and (2.b) do not cause any gain updates, while the case (1.a) causes decrease in $T_i$'s gain by an amount of $x.size$ (lines 3-5 in Alg.~\ref{alg:process-child}) and the case (2.a) causes an increase in $T_i$'s gain by an amount of $x.size$ (lines 10-12 in Alg.~\ref{alg:process-child}).

Apart from the gain updates, Alg.~\ref{alg:process-child} may also update functions $\tau$ and $\delta$ as well as $T_i$'s predecessor information.
%
%These are relatively straightforward.
%
At the end of Alg.~\ref{alg:process-child} $x$'s outgoing neighbors with state \texttt{AVAIL} excludes $u$ as $u$'s state is not \texttt{AVAIL} any more (line 14).
If nothing depends on $x$, we can release the tensor represented by $x$ from memory (line 15-16).

%% PAR process node 1~2 paragraphs
\emph{\textbf{Updates due to node $u$ being processed in the selected contraction tree.}}
After the gain updates due to $u$'s children have been completed (lines 3-4 in Alg.~\ref{alg:process-ctree}), the final task is to perform the updates due to node $u$'s state change from \texttt{AVAIL} to either \texttt{INMEM} or \texttt{RELEASED}.
These are presented in Alg.~\ref{alg:process-node} and consist of gain updates stemming from individual and coarse gains as well as setting of functions $\tau$ and $\delta$ for $u$.
Note that we process the nodes in $T^{\prime}$, of which $u$ is one, but $u$ may also appear in other trees.
After processing $u$, it will not be an element of these trees, hence we exclude the individual gains contributed by $u$ (lines 1-2).
Then, we set up functions $\tau$ and $\delta$ for each tree $u$ was in (lines 3-12).
For this purpose we visit the trees for which $u$ has at least one outgoing edge, i.e., $\bigcup_{v \in u.outAv} {v.ctree}$.
Contraction of a tree $T_i$ of which $u$ is a predecessor would cause release of the tensor represented by $u$ from memory if $u$'s all outgoing connections are only in $T_i$, which is when $\delta(u,T_i)=0$.
To find such trees, we go over all the trees $u$ is a predecessor of and update the respective gains due to coarse gains (lines 13-16).
We finally check whether we can release the tensor represented by $u$ (lines 17-20), which happens in the case $u$ has no outgoing vertex neighbors whose state is \texttt{AVAIL}, i.e., when $u.outAv$ is empty.

%% PROCESS-NODE
\begin{algorithm}[t]
\caption{PROCESS-NODE ($u$)}
\label{alg:process-node}
\begin{algorithmic}[1]
\footnotesize

\Statex // Individual gain updates
\FOR{$T_i \in u.ctree$}
  \State $T_i.tgain = T_i.tgain - T_i.u.igain$ \vskip 4pt
\ENDFOR

\Statex // Setup functions $\tau(u, \cdot)$ and $\delta(u, \cdot)$
\State $S = \emptyset$
\FOR{$v \in u.outAv$}
  \FOR{$T_i \in v.ctree$}
    \IF{$T_i \notin S$}
      \State $\tau(u, T_i) = 0$
      \State $\delta(u, T_i) = |u.outAv|$
      \State $S = S \cup T_i$
      \State $T_i.pred = T_i.pred \cup u$
    \ENDIF
  \State $\delta(u, T_i) = \delta(u, T_i) - 1$
  \State $\tau(u, T_i) = \tau(u, T_i) + 1$ \vskip 4pt
  \ENDFOR
\ENDFOR

\Statex // Coarse gain updates
\FOR{$T_i \in S$}
  \IF{$\delta(u,T_i) == 0$}
    \State $T_i.cgain = T_i.cgain + u.size$
    \State $T_i.tgain = T_i.tgain + u.size$ \vskip 4pt
  \ENDIF
\ENDFOR

\IF{$u.outAv == \emptyset$}
  \State $u.state =$ \texttt{RELEASED}
\ELSE
  \State $u.state =$ \texttt{INMMEM}
\ENDIF
    
\end{algorithmic}
\end{algorithm}

% %% PAR structure for maintaining gains
% \emph{\textbf{Maintaining and selecting gains.}}
% In order to select a tree with maximum $T_i.tgain$ efficiently (line 3 in Alg.~\ref{alg:tree-sch}), we need to maintain $T_i.tgain$ values in a data structure that supports changing these values when we process a tree since the processed tree may cause updates to $T_i.tgain$ values of other trees.
% %
% The gains of a tree may be updated in different locations: line 24 in Alg.~\ref{alg:tr-init}; lines 5, 12 in Alg.~\ref{alg:process-child}; lines 2, 16 in Alg.~\ref{alg:process-node}.
% %
% The simplest way to select such a tree is to iterate over all trees that are yet to be processed and find a tree with the largest gain.
% %
% Other possible ways to maintain gains are priority queues and bucket lists.
% %
% In a priority queue, the trees would be keyed according to $T_i.tgain$.
% %
% In a bucket list, we would have a list of buckets where each bucket corresponds to a different $T_i.tgain$ and the trees with the same gain values would be maintained in the same bucket.
% %
% Bucket list is more efficient when there is not much variance among the gain values and they are close to each other whereas priority queue is better for general cases where there are no such assumptions.
% %
% We prefer the former as in our case contraction trees consist of only a few nodes and the sizes of the nodes are often the same or there are only a few nodes with distinct sizes.

\emph{\textbf{Runtime complexity.}}
%% PAR vertex and edge factors
We define $F_v$ and $F_e$ to aid the runtime complexity analysis.
$F_v$ is defined as the average number of times a vertex appears in a contraction tree, i.e., $(\sum_{u\in V}|u.ctree|)/|V|$.
Similarly, $F_e$ is defined as the average number of times an edge appears in a contraction tree, i.e., $(\sum_{(u,v)\in E}|\{T_i:u\in T_i \mbox{ and } v \in T_i\}|)/|E|$.
Although in theory $F_v$ and $F_e$ can be as large as number of all contraction trees, $k$, in practice it is constant (see Table~\ref{tb:corr-func}).

%% PAR TR-INIT
The routine TR-INIT takes $O(F_v(V+E))$ since for the source node of an edge in $E$, it traverses all the trees that source node is in, which on the average is $F_v$.
%% PROCESS-CHILD
We next analyze on the runtime complexity of all PROCESS-CHILD calls.
Observe that this routine is called for each edge in $E$ and not for each replicated edge in trees.
When it is called for an edge $(x,u)$ it traverses the successor trees of $x$ and makes necessary updates, each of which takes constant time.
Hence, it takes $O(kE)$.
%% PROCESS-NODE
We next focus on the runtime complexity of all PROCESS-NODE calls.
Observe that PROCESS-NODE is called once for each node, and once it is called for node $u$, $u$'s outgoing edges are checked in line 4 of Alg.~\ref{alg:process-node}.
Hence, each edge in $E$ is checked once in all calls.
When making this check for edge $(u,v)$, each tree in which the target node $v$ appears needs to be processed (line 5 of Alg.~\ref{alg:process-node}). 
The number of such trees is on the average $F_v$.
Hence, all PROCESS-NODE calls take a total of $O(F_vE)$.

As a result, tree scheduler takes a total of $O(F_vV + F_vE + kE)$.
Since, $F_v \leq k $ and in practice $|E| > |V|$, resulting in $O(kE)$.

\begin{table}[t]
  \caption{Properties of the correlation functions.}
  \vspace{-3ex}
  \begin{center}
    \scalebox{0.68} {
      \begin{tabular}{l r r r r r r r r r}
        \toprule
        & & & & & \multicolumn{4}{c}{Contraction DAG}  \\
          \cmidrule(lr){6-9} 
        Corr. Func. & Type & \#Trees & Complexity & $N$ & \#Vertices & \#Edges & $F_v$ & $F_e$  \\
        \midrule
	\texttt{a0-111} & \textit{MxM} & 19041 & $O(N^3)$ & 1024 &  18552 & 36120 & 5.09 & 4.09   \\
        \texttt{a0-d3} & \textit{MxM} & 3921 & $O(N^3)$ & 1536 & 3826 & 7232 & 4.83 & 3.83  \\
        \texttt{f0} & \textit{MxMxM} & 27999 & $O(N^3)$ & 768 & 30473 & 59416 & 4.95 & 3.96   \\
        \texttt{roper} & \textit{BxM} & 84894 & $O(N^4)$ & 64 & 90378 & 180008 & 5.67 & 4.67   \\
        \texttt{deuteron} & \textit{BxB} & 109444 & $O(N^4)$ & 64 & 156508 & 312720 & 7.00 & 6.00  \\
        \texttt{tritium} & \textit{BxBxB} & 6085 & $O(N^5)$ & 32 & 7597 & 15178 & 10.11 & 9.75   \\
	  \bottomrule
        \end{tabular}
        } 
  \end{center}
  \label{tb:corr-func}  
\end{table}

\section{Experiments}
\label{sec:exp}

\subsection{Setting}
%% PAR correlation functions
We conduct our experiments on six correlation functions consisting of different meson and baryon configurations.
For these function computations we construct the contraction DAG as described in Section~\ref{sec:pre}.
As it is challenging to evaluate the correlation functions for larger distillation space basis $N$ and Redstar can run out of memory for large $N$, we select a large enough value for $N$ that does not go out of memory.
We test hadron nodes with different $N$ for our experiments.
The properties of the tested correlation functions are presented in Table~\ref{tb:corr-func}.
The columns on the left of the table summarize the properties of the correlation functions while the columns under ``Contraction DAG'' summarize the properties of the formed contraction DAG for the computation of these functions.

%% PAR systems
We evaluate the schedulers on a system that consists of a NVIDIA A100 GPU (Ampere architecture), which has 40 GB HBM2 memory with 1215 MHz clock rate and 6912 cores with 1.41 GHz clock rate, and a 64-core AMD EPYC 7763 (Milan) CPU with 2.45 GHz clock rate and a total of 256 GB DDR4 memory.
The CPU and GPU is connected with PCIe 4.0.
The host codes are compiled with GNU gcc 13.2 and the device codes use CUDA Toolkit version 12.4.
We integrated our schedulers into Redstar software suite, specifically the component that is responsible for managing correlation function computations.
%
% Our schedulers can be found at \url{https://github.com/JeffersonLab/redstar}.
%
% The values for $N$ used for each GPU system are presented in Table~\ref{tb:gpu-vecsize}.

% \begin{table}[h]
%   \caption{Distillation space basis $N$ (i.e., matrix or tensor dimension) used on GPU systems.}
%   % \vspace{-2ex}
%   \begin{center}
%     \scalebox{1.00} {
%       \begin{tabular}{l r r r}
%         \toprule
%         Corr. Func. & A100/40GB & V100/16GB & A6000/48GB  \\
%         \midrule
% 	\texttt{a0-111} & 1024 &  &   \\
%         \texttt{a0-d3} & 1536 &  &   \\
%         \texttt{f0} & 768 &  &   \\
%         \texttt{roper} & 64 &  &   \\
%         \texttt{deuteron} & 64 &  &   \\
%         \texttt{tritium} & 32 &  &   \\
% 	  \bottomrule
%         \end{tabular}
%         } 
%   \end{center}
%   \label{tb:gpu-vecsize}  
% \end{table} 

\begin{figure}[t]
  \centering
  \subfloat[\texttt{a0-111}]{\includegraphics[width=0.10\textwidth]{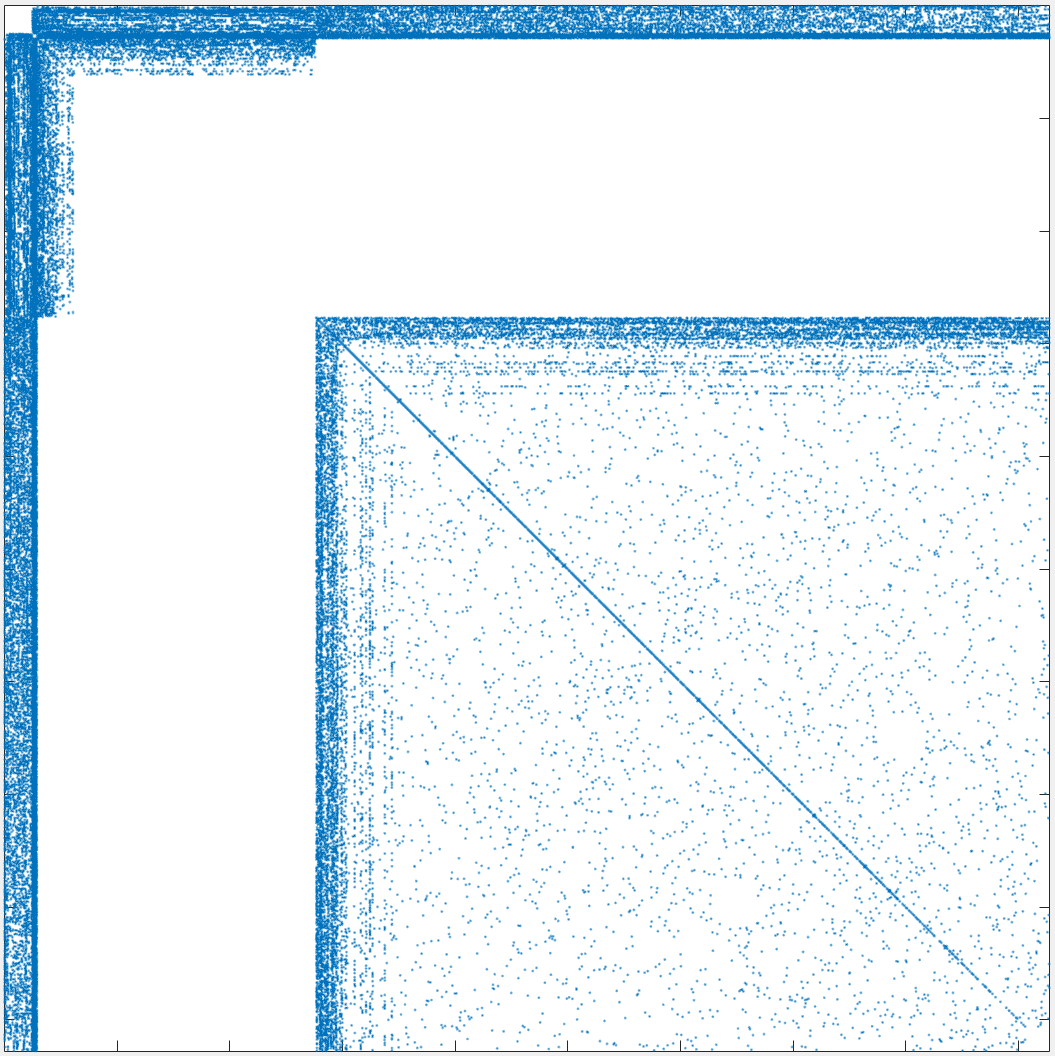} } \quad 
  \subfloat[\texttt{a0-d3}]{\includegraphics[width=0.10\textwidth]{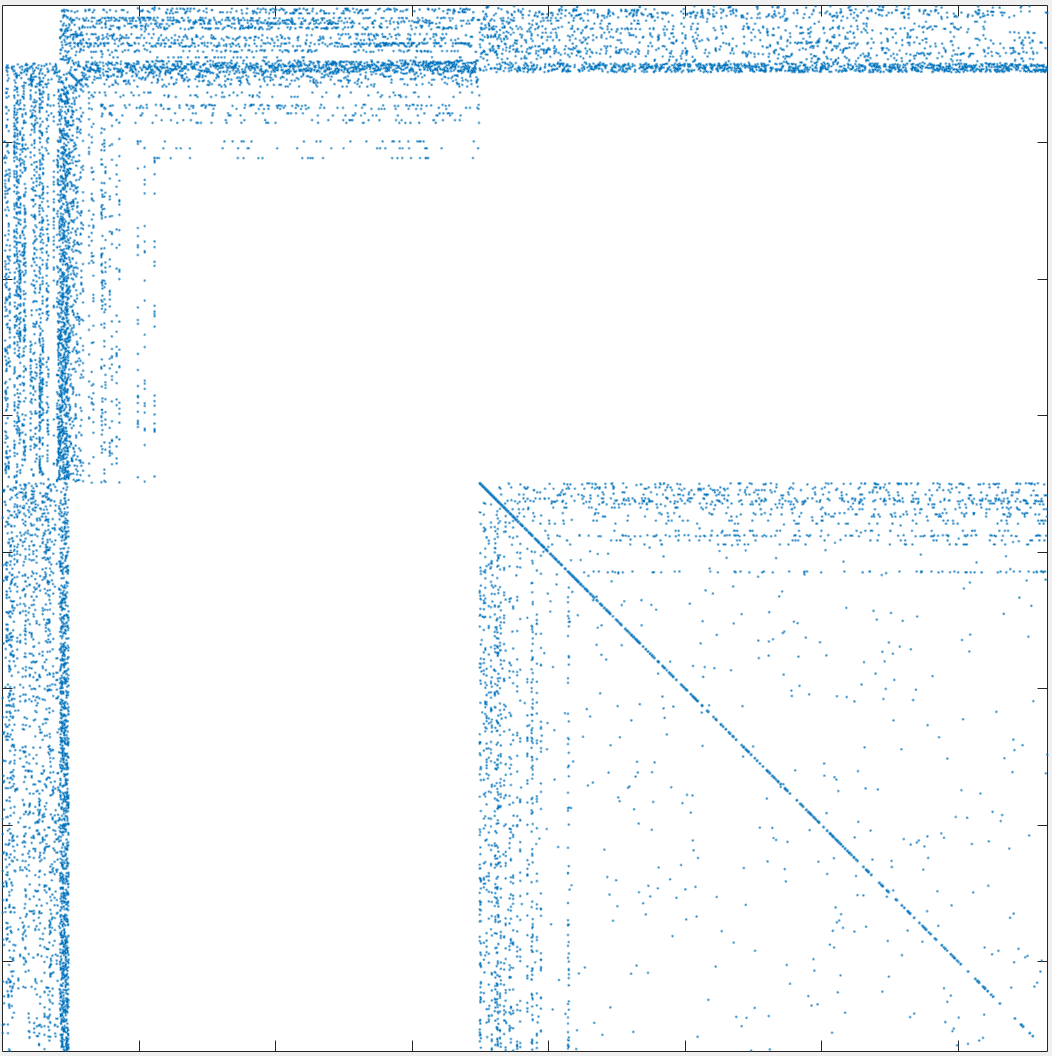} } \quad 
  \subfloat[\texttt{f0}]{\includegraphics[width=0.10\textwidth]{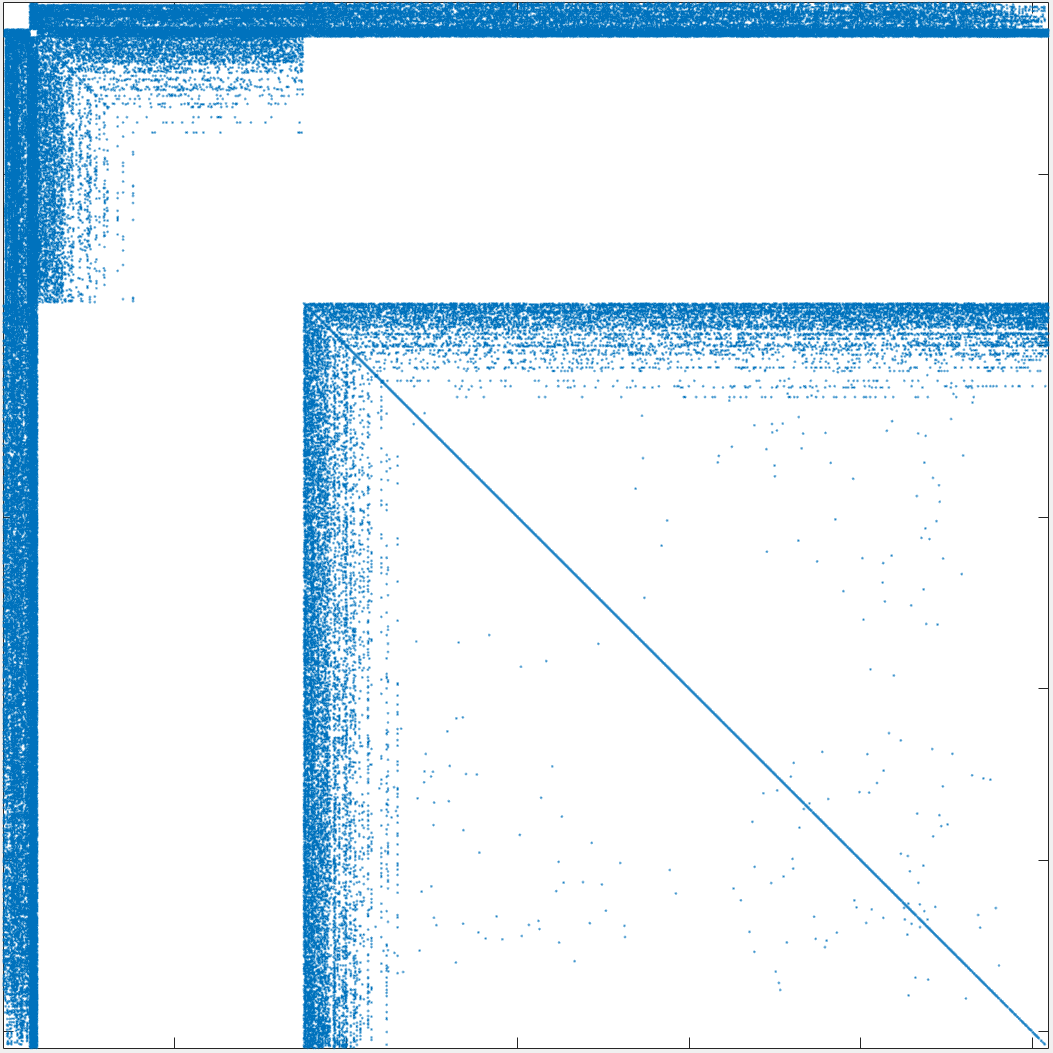} } \\ 
  \subfloat[\texttt{roper}]{\includegraphics[width=0.10\textwidth]{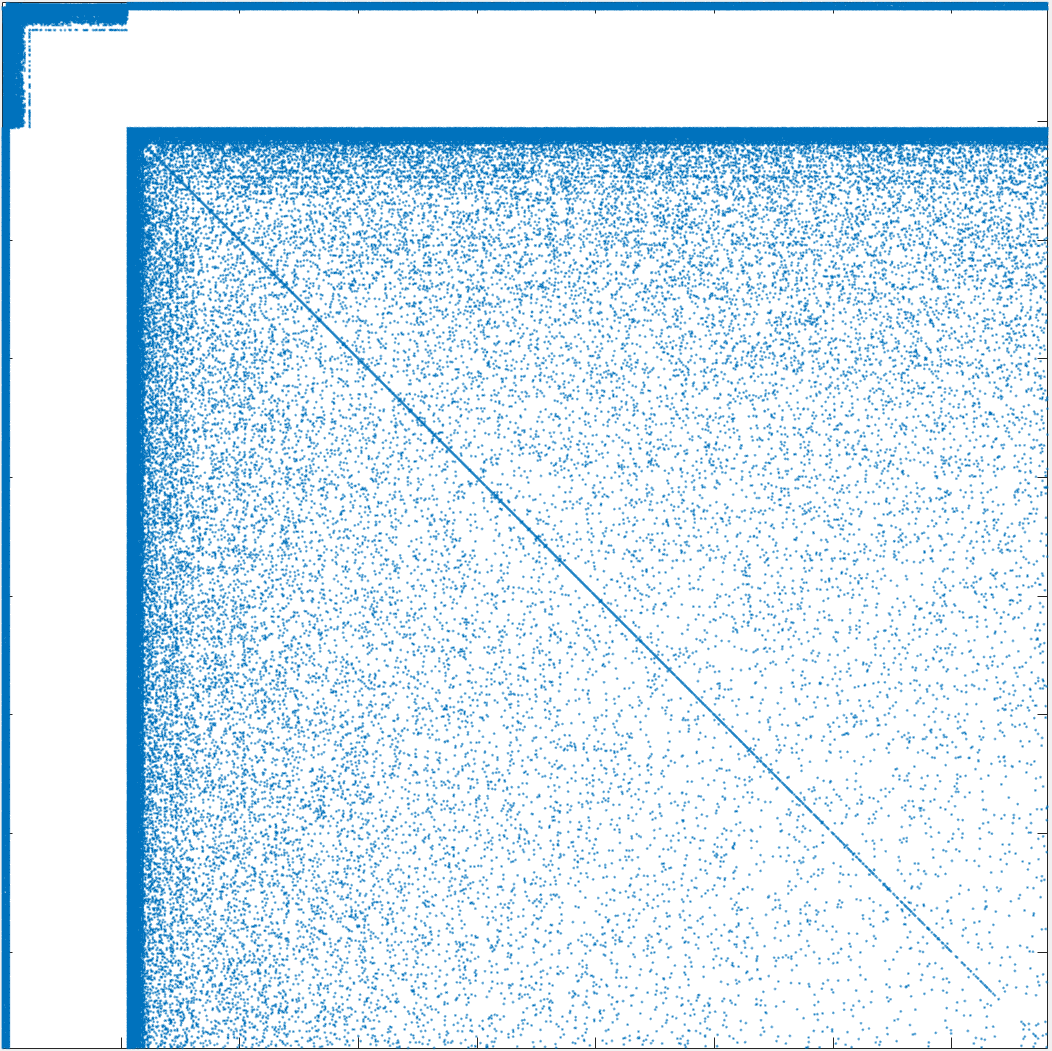} } \quad 
  \subfloat[\texttt{deuteron}]{\includegraphics[width=0.10\textwidth]{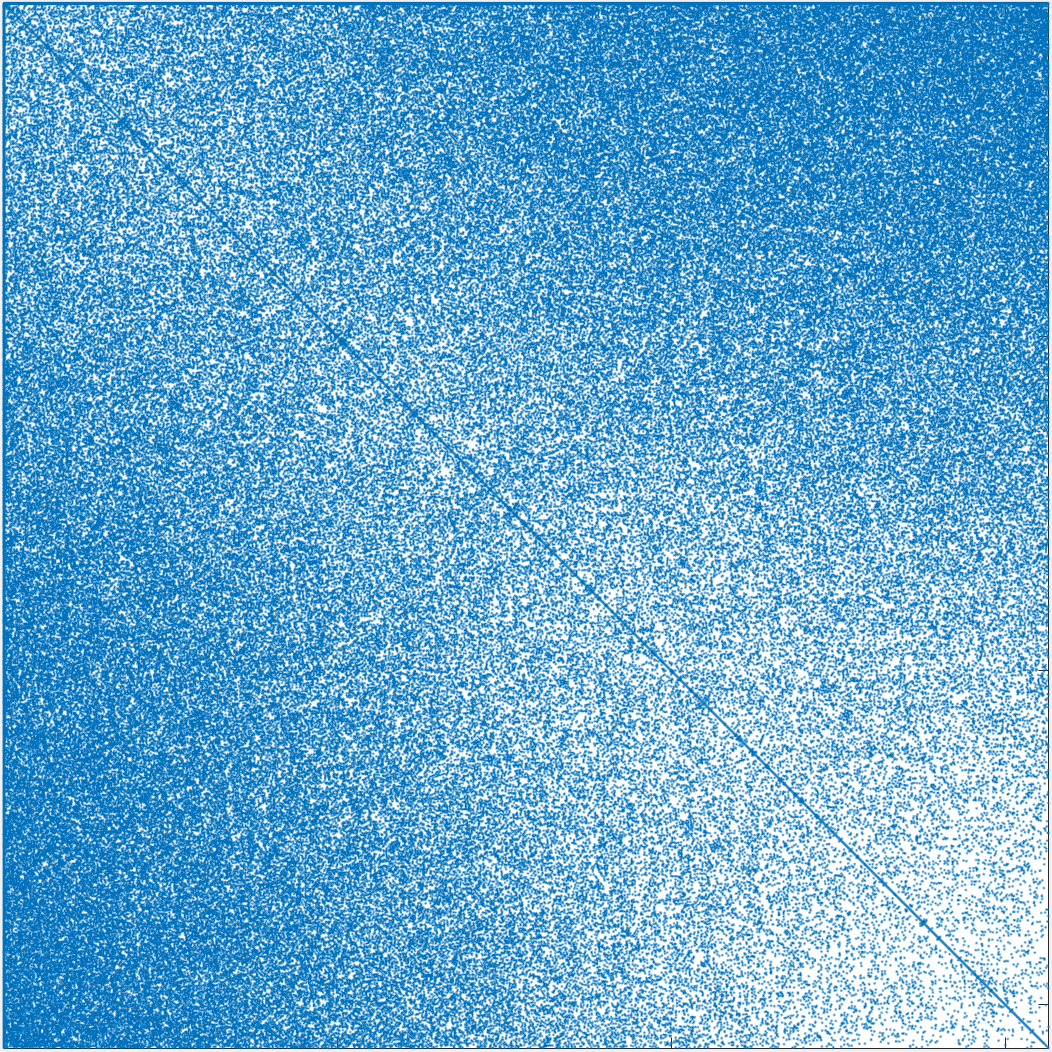} } \quad 
  \subfloat[\texttt{tritium}]{\includegraphics[width=0.10\textwidth]{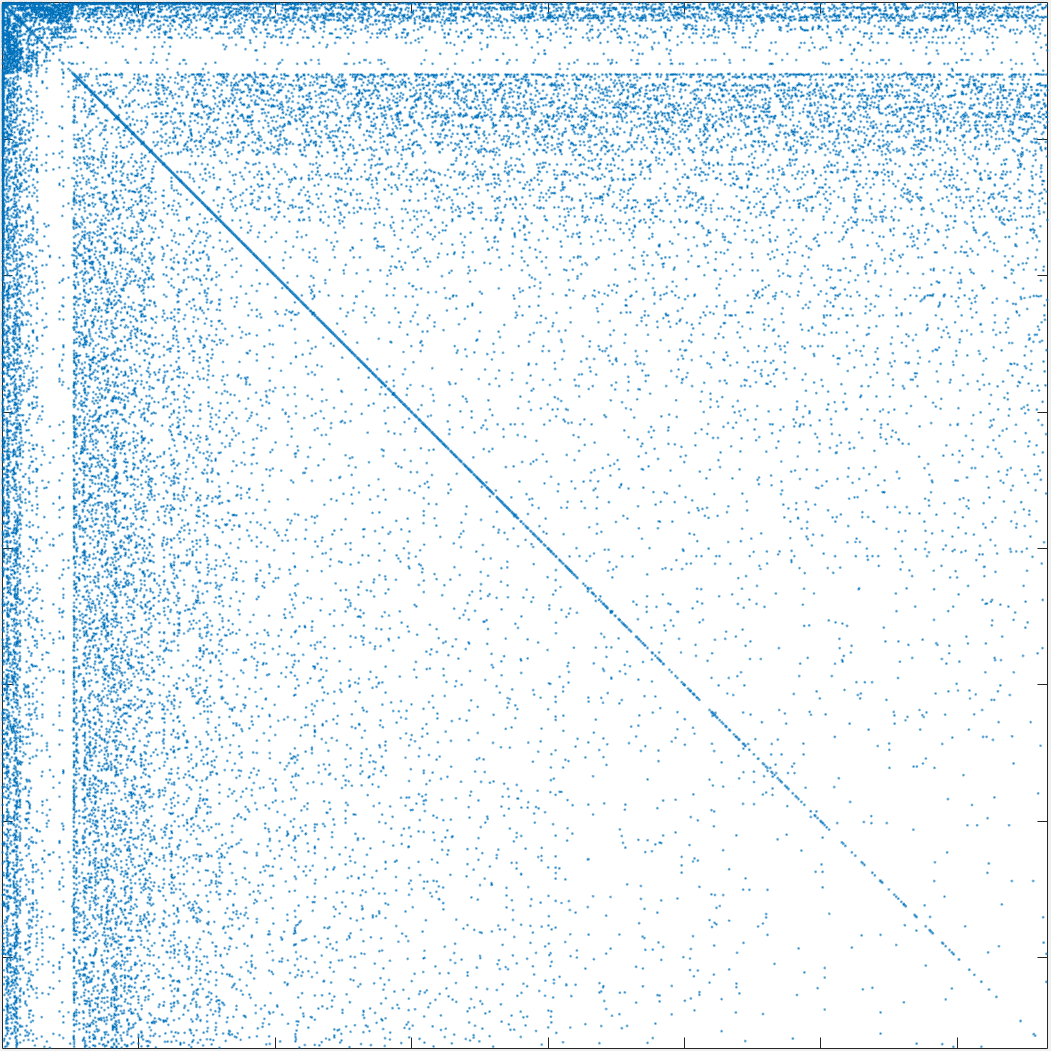} } 
  \caption{Structures of the contraction DAGs.}
  \label{fig:cdag-pattern}
  % \vspace{-3ex}
\end{figure}

\subsection{Assessment of peak memory on the contraction DAGs}
\label{sec:cdag-peak}
\begin{figure*}[t]
  \centering
  \subfloat{\includegraphics[scale=0.24]{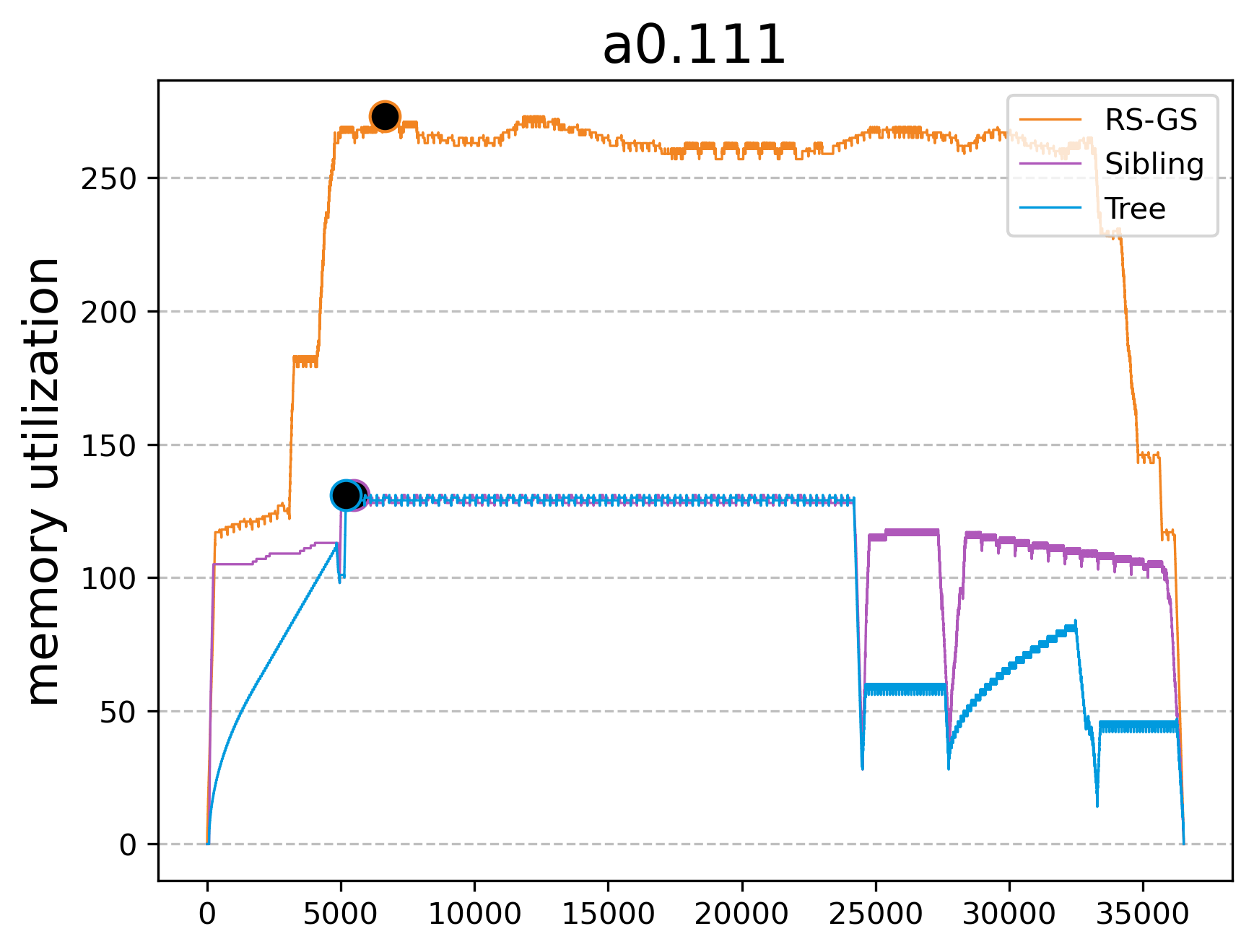}} \quad  
  \subfloat{\includegraphics[scale=0.24]{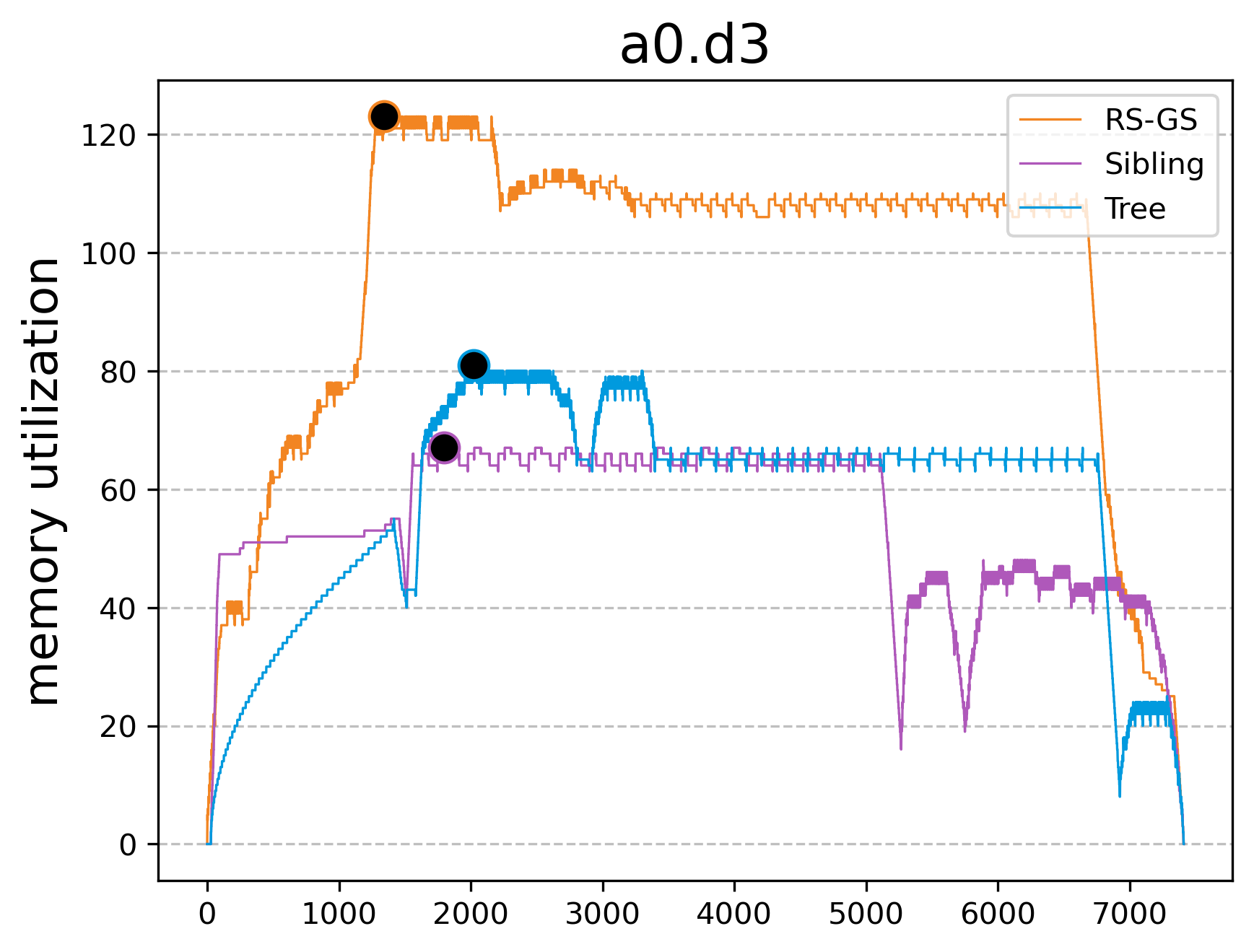}} \quad
  \subfloat{\includegraphics[scale=0.24]{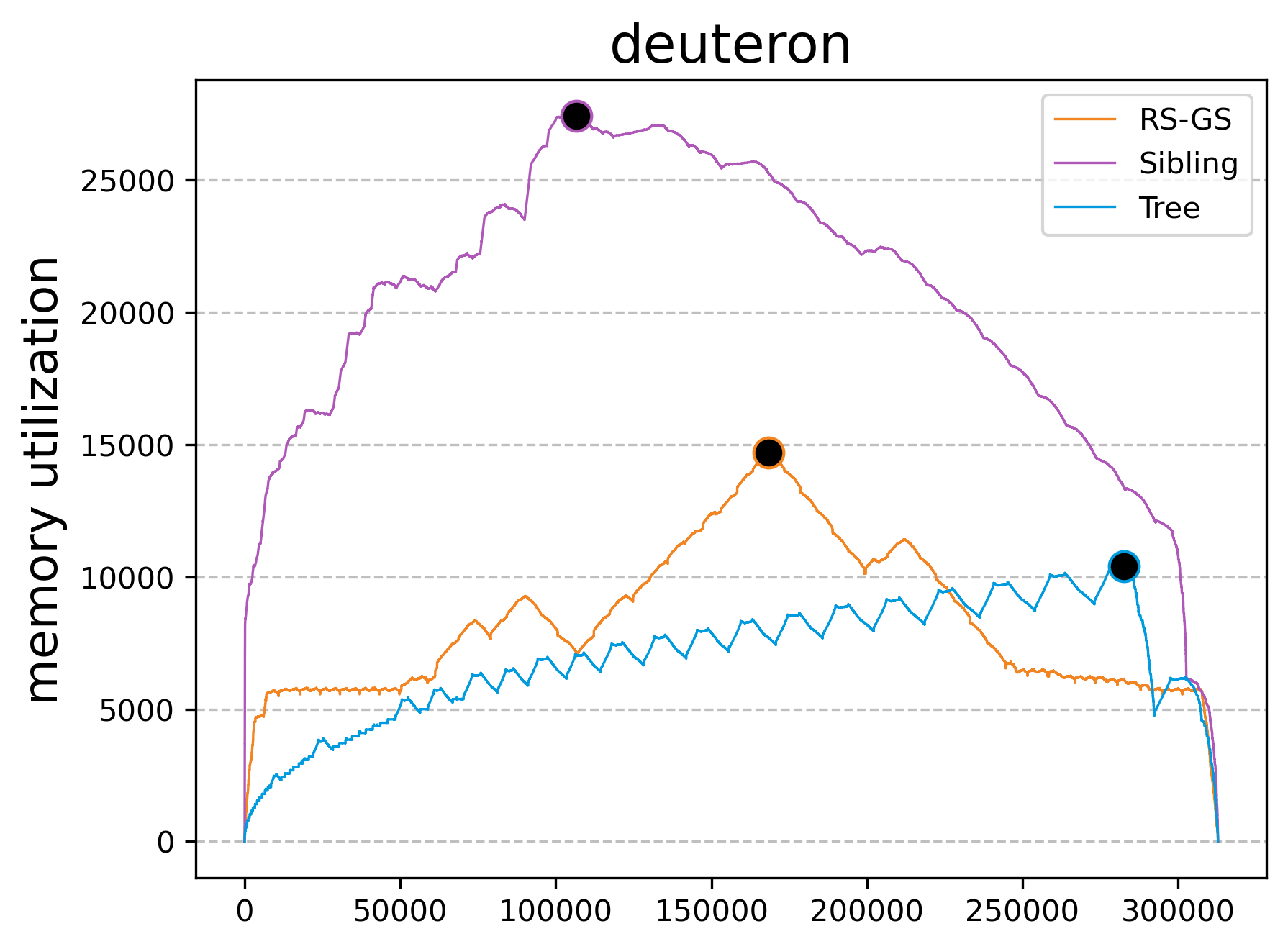}} \\
  \subfloat{\includegraphics[scale=0.24]{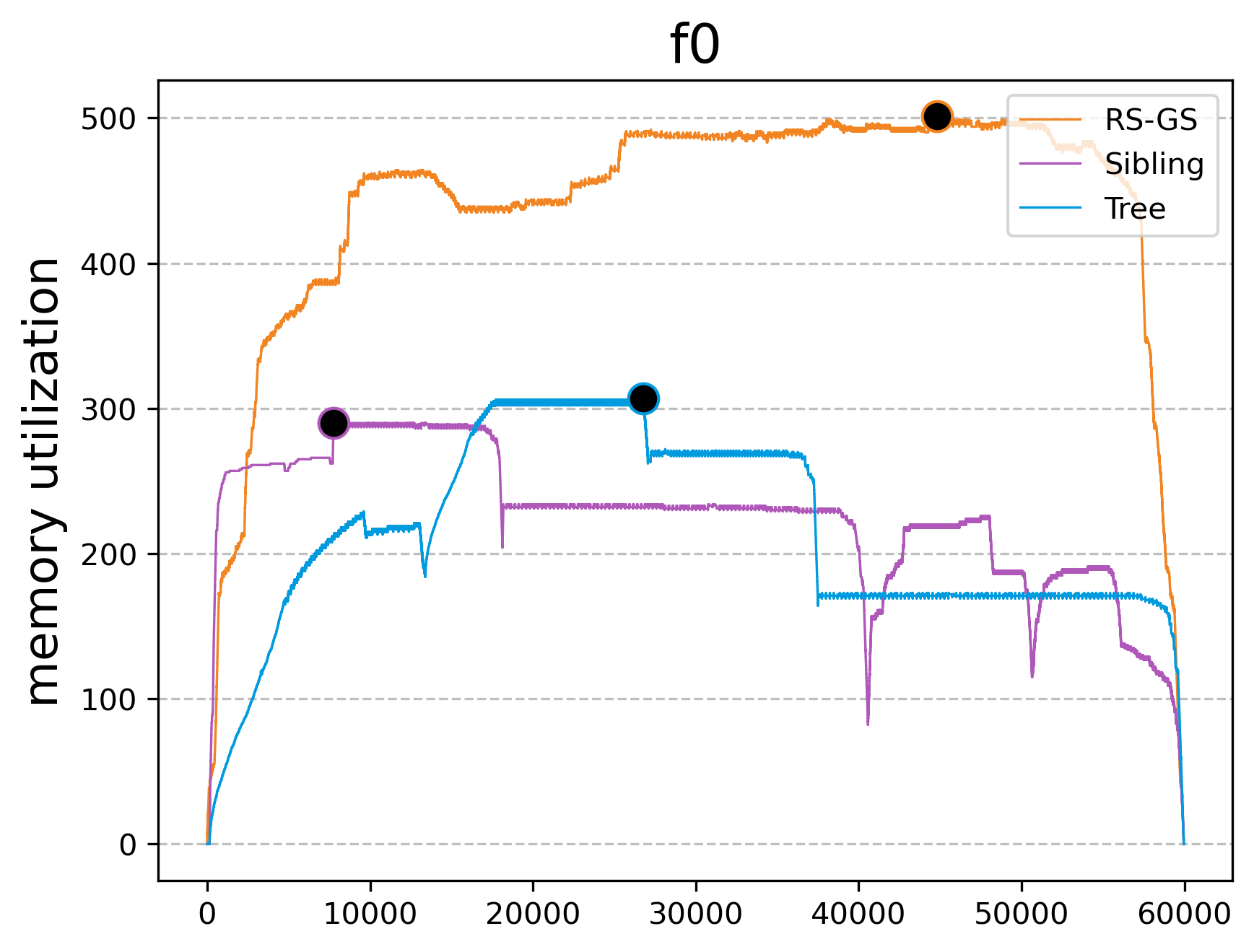}} \quad
  \subfloat{\includegraphics[scale=0.24]{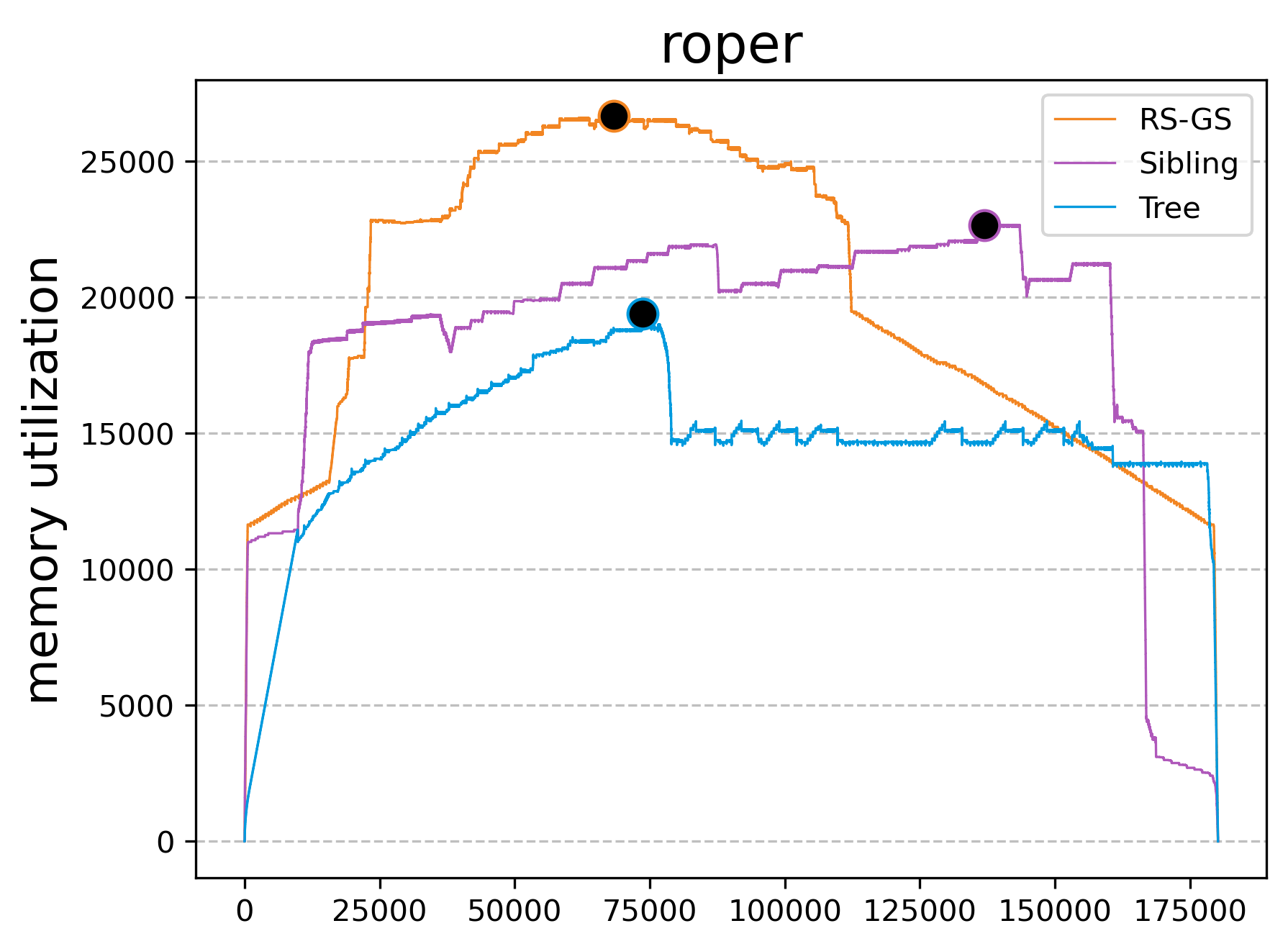}} \quad
  \subfloat{\includegraphics[scale=0.24]{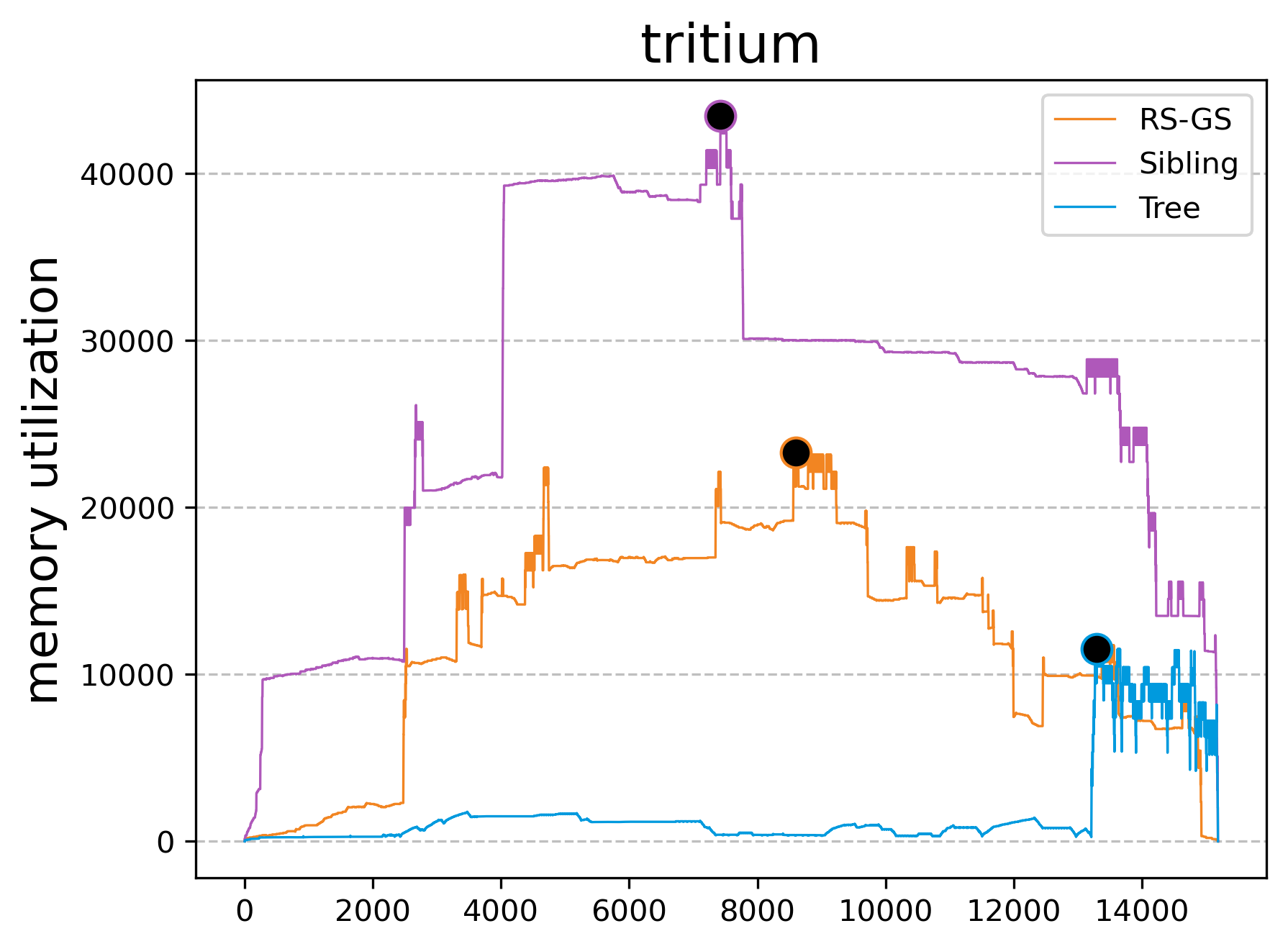}} \quad
  \caption{Memory utilization of the schedulers for the six different correlation functions. Black dots indicate the peak memory.}
  \label{fig:peakmem}
  \vspace{-3ex}
\end{figure*}

%% PAR simple exec queues
% how is peak memory computed?
We compare the peak memory obtained by the schedulers in terms of the metric described in Section~\ref{sec:pre-problem} on the contraction DAGs.
The purpose of this section is to assess the effectiveness of the schedulers in terms of the objective.
To do so, we utilize execution queues produced in Redstar using the schedulers.
Each operation in this queue may be one of three operations: (i) exterior contract for interior nodes, (ii) contract all for root nodes, and (iii) tensor deletion.
The tensors corresponding to the leaf nodes and that are required by a contraction are brought to memory if they are not in memory.
After each operation in the queue, we record the total sum of sizes of the nodes that are in memory.
The nodes sizes are divided by the greatest common divisor of all distinct sizes to make the plots more presentable.
Recall that node sizes correspond to sizes of the tensors they represent, which may be one of $\{O(N^2), O(N^3), O(N^4)\}$.
Not all of them appear in a specific correlation function: in \texttt{a0-111} (a \textit{MxM} system) only $O(N^2)$ sizes exist while in \texttt{tritium} (a \textit{BxBxB} system) all of them exist.
We plot the memory utilization obtained by the schedulers in Fig.~\ref{fig:peakmem}, where the $x$ axis is the operation id in the queue and $y$ axis is the   memory occupied after that operation.
The compared schedulers are Redstar's graph sorting ordering of the contractions, abbreviated as \texttt{RS-GS} (Section~\ref{sec:pre}), sibling scheduler, abbreviated as \texttt{Sibling} (Section~\ref{sec:sibling}), and tree scheduler, abbreviated as \texttt{Tree} (Section~\ref{sec:tree}).

%% PAR results
When we compare the schedulers in the plots in Fig.~\ref{fig:peakmem}, it is seen that the proposed schedulers do a good job of addressing the target metric compared to \texttt{RS-GS}.
In all six datasets, one of the two proposed schedulers achieves smaller peak memory than \texttt{RS-GS}.
In datasets \texttt{a0-111}, \texttt{a0-d3}, \texttt{f0}, \texttt{roper}, \texttt{deuteron}, and \texttt{tritium}, the best of the \texttt{Sibling} and \texttt{Tree} attains 2.1x, 1.8x, 1.7x, 1.4x, 1.4x, and 2.0x peak memory improvement over \texttt{RS-GS}, respectively.
In relatively smaller instances \texttt{a0-111}, \texttt{a0-d3}, and \texttt{f0}, \texttt{Sibling} performs the best, though \texttt{Tree} attains close values to it.
In the larger instances \texttt{roper}, \texttt{deuteron}, and \texttt{tritium}, \texttt{Tree} is clearly the best performing scheduler, while \texttt{Sibling}'s performance drops below \texttt{RS-GS} in two of the instances.
The worse performance of \texttt{Sibling} in larger instances can be attributed to its disregard of the node sizes and locality of the scheduling decisions.
\texttt{Tree} on the other hand, attains superior peak memory, which validates the design motivation of it: having global view of the state of the memory and making scheduling decisions dynamically based on this information results in better memory optimization.
Overall, it can be said that \texttt{Tree} is preferable over \texttt{Sibling} and \texttt{RS-GS}.

\subsection{Effect of scheduling algorithms in Redstar}
%% PAR Integration into redstar, metrics tested
In this section we evaluate the performance of the schedulers in the LQCD analysis software Redstar.
After forming the contraction DAG in Redstar, prior to performing the contractions we run schedulers to determine an order.
From this order, an execution queue is produced and it is used to perform contractions or remove tensors that are not needed as the operations take place.
We use four different metrics to assess the schedulers: (i) overall time spent in contractions, (ii) total number of evictions, (iii) total number of data transfers between host and device, and (iv) total size of data transfers between host and device.
Recall that an eviction happens when a contraction is to be performed on the device but there is not enough device memory for performing that contraction.
We present the time spent for the contractions, which consists of allocation, transfer, and deletion of tensors as well as performing the contraction.
Hence, it is a direct measure of the effectiveness of the schedulers within the application.
There are other components of the Redstar, but those are not of interest to our work, and contractions constitute a big portion of the overall application time.
We also present more detailed metrics number and size of transfer operations between host and device to assess the reduction in data movement.
Among these four metrics, reducing the number of evictions is a direct consequence of reducing the peak memory (which is the goal of the proposed schedulers), and the remaining metrics follow from it.

\begin{figure*}[t]
  \centering
  \subfloat{\includegraphics[width=0.22\textwidth]{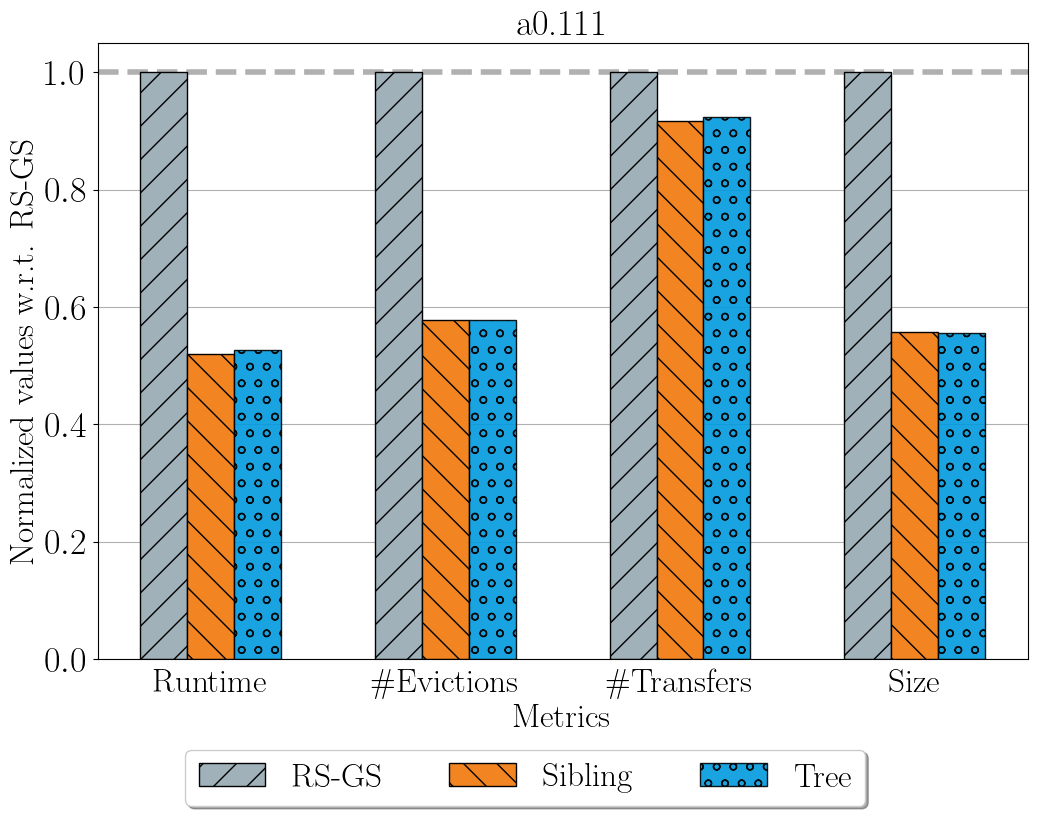}} \quad  
  \subfloat{\includegraphics[width=0.22\textwidth]{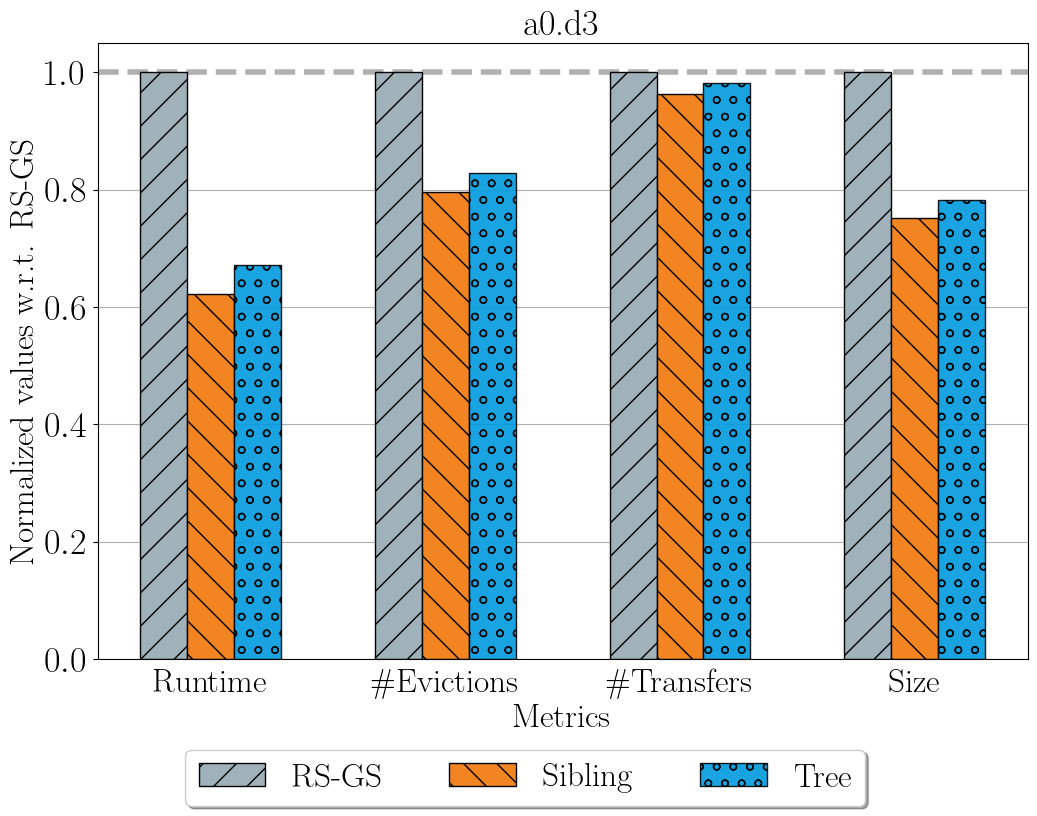}} \quad
  \subfloat{\includegraphics[width=0.22\textwidth]{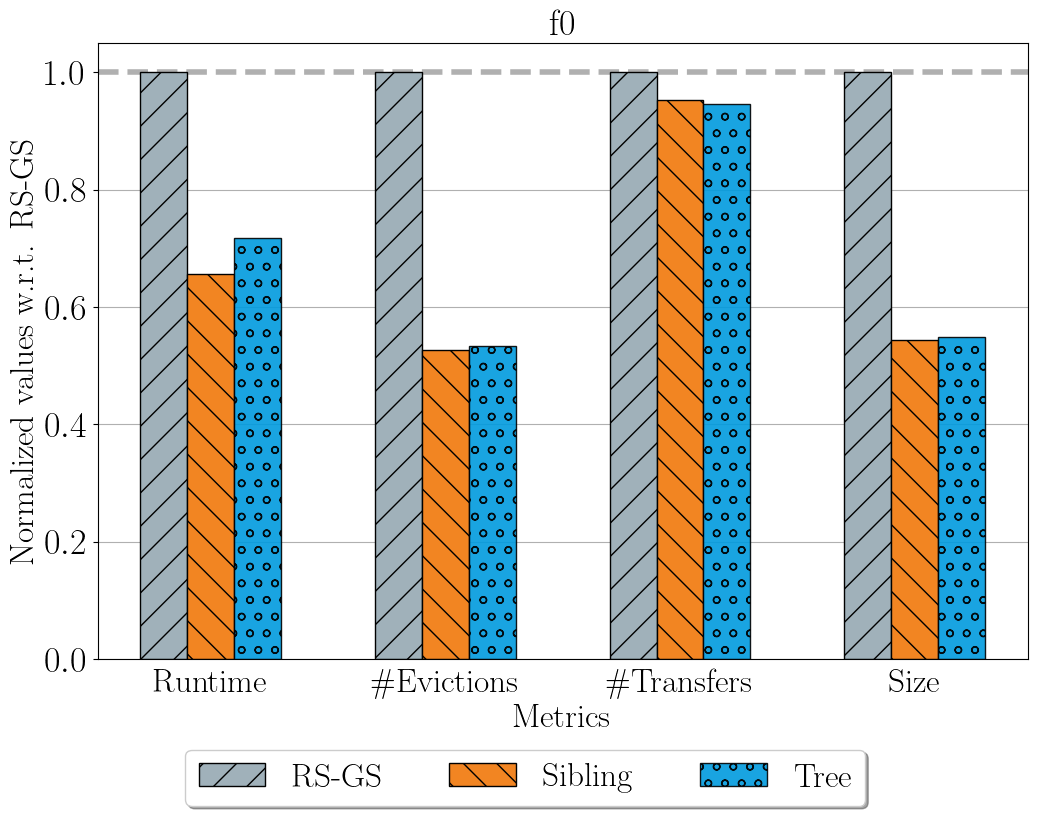}} \\
  \subfloat{\includegraphics[width=0.22\textwidth]{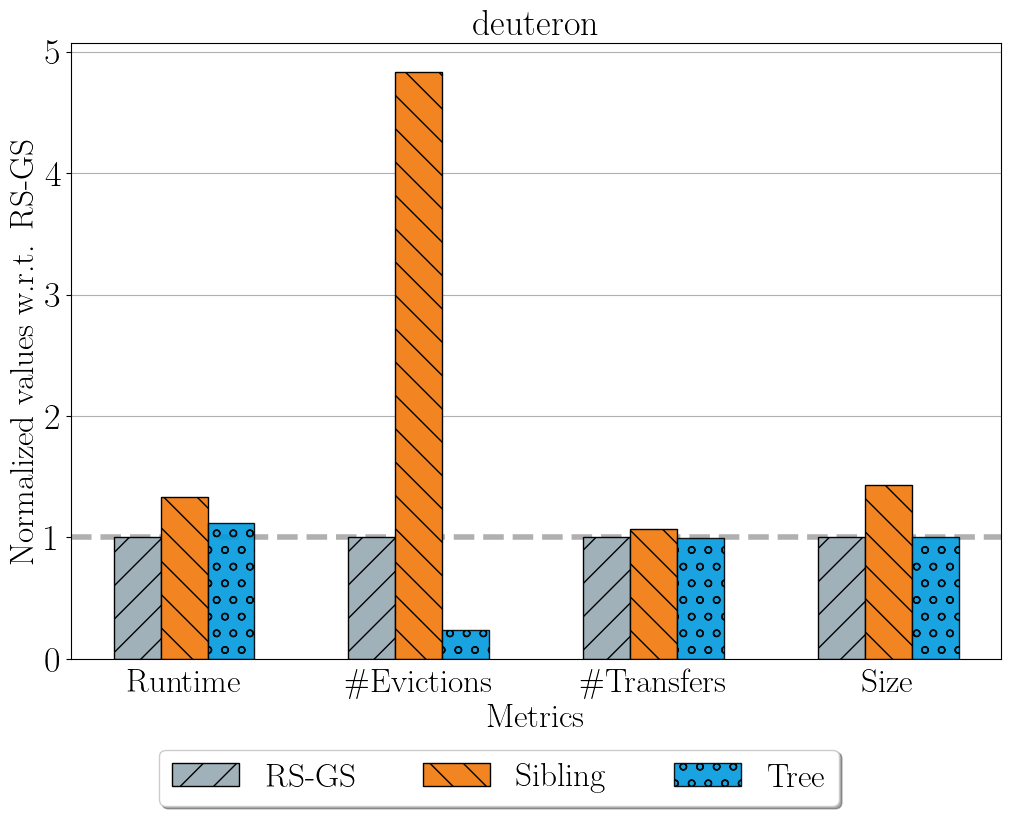}} \quad
  \subfloat{\includegraphics[width=0.22\textwidth]{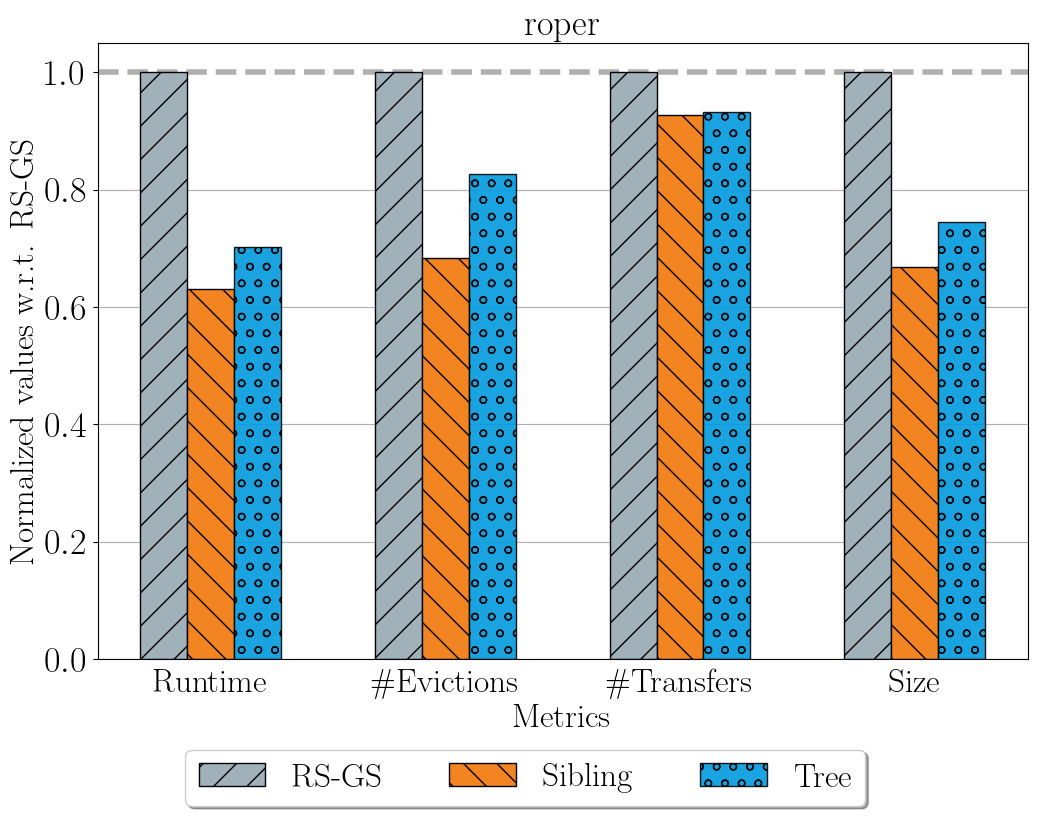}} \quad
  \subfloat{\includegraphics[width=0.22\textwidth]{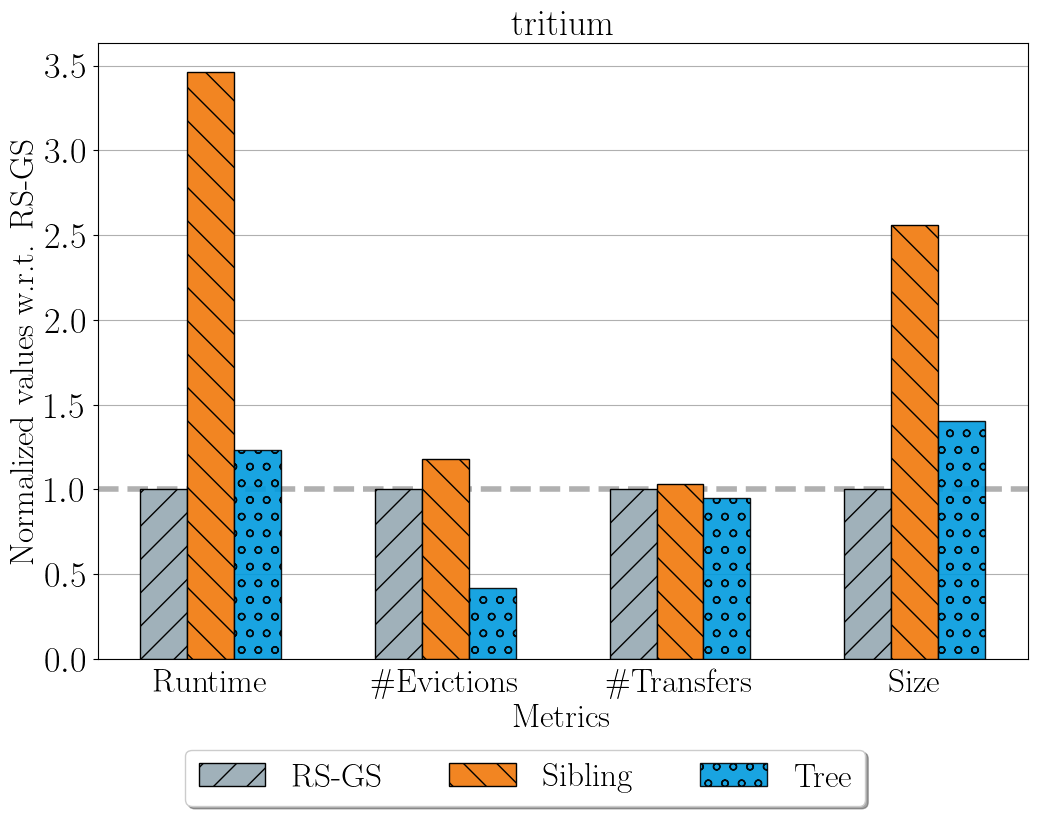}} \quad
  \caption{Comparison of schedulers in terms of four different metrics within the LQCD analysis software Redstar on six different correlation functions. The bars below 1.0 signify an improvement over the baseline scheduler \texttt{RS-GS}.}
  \label{fig:redstar-bars}
  \vspace{-3ex}
\end{figure*}

%% PAR explain the figure
We present the values obtained by the three schedulers regarding the four described metrics on six correlation functions in Fig.~\ref{fig:redstar-bars}.
The four metrics time spent in contractions, number of evictions, number of data transfers, and size of data transfers described above are presented in charts as ``Runtime'', ``\#Evictions'', ``\#Transfers'', and ``Size'', respectively.
The values obtained by \texttt{Sibling} and \texttt{Tree} in each metric are normalized with respect to the values obtained by \texttt{RS-GS} at that metric.
For the bars in the figures, a bar lower than 1.0 indicates better performance in that metric.
For example, a value 0.5 for a bar in metric ``Size'' indicates a 2x reduction in amount of data transferred over \texttt{RS-GS}.

%% PAR #evictions
When we compare the number of evictions attained by the schedulers in Fig.~\ref{fig:redstar-bars}, it can be seen that in all instances \texttt{Sibling} and \texttt{Tree} lead to significant reductions.
In smaller three datasets (\texttt{a0-111}, \texttt{a0-d3}, \texttt{f0}), \texttt{Sibling} performs slightly better than \texttt{Tree} in this metric and as the datasets get bigger \texttt{Tree} gets to perform the best.
Observe that the values attained by the schedulers are in agreement with our assessment in Section~\ref{sec:cdag-peak}, validating the effectiveness of the target objective of minimizing peak memory utilization.
When we consider the best of the two proposed schedulers, the improvements over \texttt{RS-GS} in the number of evictions are  1.7x, 1.3x, 1.9x, 1.5x, 4.2x, and 2.4x for \texttt{a0-111}, \texttt{a0-d3}, \texttt{f0}, \texttt{roper}, \texttt{deuteron}, and \texttt{tritium}, respectively.
The effectiveness of \texttt{Tree} is especially noteworthy in larger datasets \texttt{deuteron} and \texttt{tritium}.

\begin{table}[t]
  \caption{Total size of data movement (TB).}
  \vspace{-4ex}
  \begin{center}
    \scalebox{0.78} {
      \begin{tabular}{l r r r}
        \toprule       
        Corr. Func. & \texttt{RS-GS} & \texttt{Sibling} & \texttt{Tree}  \\
        \midrule
	\texttt{a0-111} & 2.00 & 1.12 & \textbf{1.11} \\
        \texttt{a0-d3} & 1.12 & \textbf{0.84} & 0.88 \\
        \texttt{f0} & 1.01 & \textbf{0.55} & \textbf{0.55} \\
        \texttt{roper} & 7.35 & \textbf{4.91} & 5.47 \\
        \texttt{deuteron} & 0.21 & 0.29 & \textbf{0.20} \\
        \texttt{tritium} & \textbf{0.53} & 1.34 & 0.74 \\
	  \bottomrule
        \end{tabular}
        } 
  \end{center}
  \label{tb:data-transfer}  
\end{table}

%% PAR number and size transfers
The proposed scheduling algorithms often reduce the number and size of data transfers, which are presented with bars ``\#Transfers'' and ``Size'' in Fig.~\ref{fig:redstar-bars}.
In \texttt{a0-111}, \texttt{a0-d3}, \texttt{f0}, and \texttt{roper}, our schedulers are able to reduce data movement on the PCIe link, while in \texttt{deuteron} they attain similar numbers with \texttt{RS-GS}, and in \texttt{tritium} they increase size of data movement despite \texttt{Tree} scheduler reduced both the number of evictions and data transfers.
We present the amount of data moved in Table~\ref{tb:data-transfer}, with the lowest entry in each row is indicated in bold.
As seen from the table, the proposed schedulers are able to reduce the amount of data moved significantly with improvements up to 1.84x.
In \texttt{deuteron} and \texttt{tritium}, the proposed schedulers are not able to make gains.
This can be attributed to hadron nodes having varying sizes, which may prevent the reductions obtained in theory to be reflected in practice.

%% PAR contraction time
We finally compare the contraction time in Redstar that results from the schedules computed by the schedulers in Fig.~\ref{fig:redstar-bars}.
As seen from the metric ``Runtime'', the improvements obtained in data transfers reflected in the execution time of the contractions.
In four of the six correlation functions, we observe significant reduction in computing correlation functions, while in the remaining two (\texttt{deuteron} and \texttt{tritium}), the proposed scheduler \texttt{Tree} is slightly worse than the baseline.
These runtime results validate the claim that our scheduler \texttt{Tree} is a safe and sound alternative for scheduling hadronic correlation function computations.

\begin{table}[t]
  \caption{Time spent in computing a schedule (msec).}
  \vspace{-4ex}
  \begin{center}
    \scalebox{0.78} {
      \begin{tabular}{l r r r}
        \toprule       
        Corr. Func. & \texttt{RS-GS} & \texttt{Sibling} & \texttt{Tree}  \\
        \midrule
	\texttt{a0-111} & 3.5 & 29.5 & 189.1 \\
        \texttt{a0-d3} & 0.5 & 3.8 & 19.2 \\
        \texttt{f0} & 5.9 & 50.0 & 295.2 \\
        \texttt{roper} & 26.0 & 234.3 & 3095.9 \\
        \texttt{deuteron} & 38.4 & 451.0 & 17005.5 \\
        \texttt{tritium} & 1.5 & 25.0 & 212.5 \\
	  \bottomrule
        \end{tabular}
        } 
  \end{center}
  \label{tb:sch-time}  
\end{table}

\subsection{Scheduling overhead}
%% PAR table
We compare the time spent by the schedulers to find a contraction order for six different systems.
Table~\ref{tb:sch-time} presents the runtime of the schedulers in milliseconds.
As seen from the table, proposed schedulers run slower than the sorting used by the Redstar for scheduling.
The higher scheduling runtime of \texttt{Tree} is in agreement with the runtime complexity of it.
Overall, the time to find a schedule is usually negligible compared to the time for performing contractions.
For example, for \texttt{roper} dataset the overall time spent for performing contractions using \texttt{Tree} 631.1 seconds, whereas the scheduler overhead is 3.1 seconds.
The overhead of \texttt{Tree} is somewhat higher for \texttt{deuteron} dataset.
However, the scheduling overhead stays the same irrespective of the tensor dimension: once a good schedule is found, it can reused in correlation function computations of any size and while the contraction time increases with increasing tensor dimension, the scheduling overhead stays the same.

\section{Related work}
\label{sec:rw}

Scheduling of task graphs under memory or I/O constraints is an old problem starting with the study of register allocation for evaluating expressions~\cite{Sethi1973}, with several variants afterwards~\cite{Gilbert1979, Boas1979, JiaWei1981}.
In some applications the task graph may be a tree and in this case the problem gets easier~\cite{Chung2011, Liu1987}.
Several memory-aware schedulers have been proposed for such cases~\cite{Agullo2016, Dubois2015}.

%% Some other recent works
The problem of minimization of peak memory has been recently studied for the dataflow graphs~\cite{Fradet2023}.
The authors focus on two variants of the dataflow graphs, task graphs and synchronous dataflow graphs, and propose transformations to apply classical algorithms such as branch and bound.
The memory model utilized in that work, shared buffer model, largely differs from the memory model used in our work.

Another work~\cite{Kayaaslan2018} that aims to reduce peak memory focuses on series-parallel task graphs and introduces a new model that is based on graphs with weighted vertices.
The authors propose a polynomial-time algorithm, however it is not feasible in practice for large graphs.

There is also another study~\cite{Marchal2019} that tackles the memory footprint of workflows represented by task graphs in order to avoid schedules that exceed available memory.
Authors' goal is to find the schedule with minimum makespan among a set of schedules that do not exceed memory.
For this purpose they extend the graph with fictitious edges to restrict the concurrent execution of memory-intensive tasks.
The described work focuses on and uses DAGs for parallel applications on shared-memory platforms, whereas in our case temporal locality and data reuse are the main concerns in finding a good schedule.

% The memory peak minimization problem of task graphs
% is a variation of the pebble game [15], which is NP-complete.
% A naive method to solve it is to generate all the linear extensions
% of the graph, but it has at least factorial complexity [12].

\section{Conclusions}
\label{sec:conc}
We proposed two schedulers to optimize tensor reuse in batch contractions produced by the correlation functions in LQCD simulations.
Our approaches aim to optimize the peak memory of the contractions and in this way reduce evictions from device memory.
We realized and validated our schedulers in the LQCD analysis software Redstar.
We demonstrated the proposed schedulers are able to reduce time-to-solution in several tested correlation functions.
In future, we aim to develop a general node-based scheduler that uses the gain concept used by the proposed tree scheduler.
We plan to apply this scheduler to the contraction DAGs as well as general DAGs to demonstrate its efficiency.
We also plan to develop partitioning models to schedule contractions for multi-GPU systems.

\bibliographystyle{IEEEtran}
\bibliography{IEEEabrv,oguzall}

\begin{thebibliography}{10}
\providecommand{\url}[1]{#1}
\csname url@samestyle\endcsname
\providecommand{\newblock}{\relax}
\providecommand{\bibinfo}[2]{#2}
\providecommand{\BIBentrySTDinterwordspacing}{\spaceskip=0pt\relax}
\providecommand{\BIBentryALTinterwordstretchfactor}{4}
\providecommand{\BIBentryALTinterwordspacing}{\spaceskip=\fontdimen2\font plus
\BIBentryALTinterwordstretchfactor\fontdimen3\font minus \fontdimen4\font\relax}
\providecommand{\BIBforeignlanguage}[2]{{%
\expandafter\ifx\csname l@#1\endcsname\relax
\typeout{** WARNING: IEEEtran.bst: No hyphenation pattern has been}%
\typeout{** loaded for the language `#1'. Using the pattern for}%
\typeout{** the default language instead.}%
\else
\language=\csname l@#1\endcsname
\fi
#2}}
\providecommand{\BIBdecl}{\relax}
\BIBdecl

\bibitem{HadronSpectrum:2009krc}
M.~Peardon, J.~Bulava, J.~Foley, C.~Morningstar, J.~Dudek, R.~G. Edwards, B.~Joo, H.-W. Lin, D.~G. Richards, and K.~J. Juge, ``{A Novel quark-field creation operator construction for hadronic physics in lattice QCD},'' \emph{Phys. Rev. D}, vol.~80, p. 054506, 2009.

\bibitem{Abdelfattah2016}
\BIBentryALTinterwordspacing
A.~Abdelfattah, M.~Baboulin, V.~Dobrev, J.~Dongarra, C.~Earl, J.~Falcou, A.~Haidar, I.~Karlin, T.~Kolev, I.~Masliah, and S.~Tomov, ``High-performance tensor contractions for gpus,'' \emph{Procedia Computer Science}, vol.~80, pp. 108--118, 2016, international Conference on Computational Science 2016, ICCS 2016, 6-8 June 2016, San Diego, California, USA. [Online]. Available: \url{https://www.sciencedirect.com/science/article/pii/S1877050916306536}
\BIBentrySTDinterwordspacing

\bibitem{Bibireata2004}
A.~Bibireata, S.~Krishnan, G.~Baumgartner, D.~Cociorva, C.-C. Lam, P.~Sadayappan, J.~Ramanujam, D.~E. Bernholdt, and V.~Choppella, ``Memory-constrained data locality optimization for tensor contractions,'' in \emph{Languages and Compilers for Parallel Computing}, L.~Rauchwerger, Ed.\hskip 1em plus 0.5em minus 0.4em\relax Berlin, Heidelberg: Springer Berlin Heidelberg, 2004, pp. 93--108.

\bibitem{Kim2018}
\BIBentryALTinterwordspacing
J.~Kim, A.~Sukumaran-Rajam, C.~Hong, A.~Panyala, R.~K. Srivastava, S.~Krishnamoorthy, and P.~Sadayappan, ``Optimizing tensor contractions in ccsd(t) for efficient execution on gpus,'' in \emph{Proceedings of the 2018 International Conference on Supercomputing}, ser. ICS '18.\hskip 1em plus 0.5em minus 0.4em\relax New York, NY, USA: Association for Computing Machinery, 2018, p. 96–106. [Online]. Available: \url{https://doi.org/10.1145/3205289.3205296}
\BIBentrySTDinterwordspacing

\bibitem{Liu2021}
\BIBentryALTinterwordspacing
J.~Liu, D.~Li, R.~Gioiosa, and J.~Li, ``Athena: high-performance sparse tensor contraction sequence on heterogeneous memory,'' ser. ICS '21.\hskip 1em plus 0.5em minus 0.4em\relax New York, NY, USA: Association for Computing Machinery, 2021, p. 190–202. [Online]. Available: \url{https://doi.org/10.1145/3447818.3460355}
\BIBentrySTDinterwordspacing

\bibitem{Shi2016}
Y.~Shi, U.~N. Niranjan, A.~Anandkumar, and C.~Cecka, ``Tensor contractions with extended blas kernels on cpu and gpu,'' in \emph{2016 IEEE 23rd International Conference on High Performance Computing (HiPC)}, 2016, pp. 193--202.

\bibitem{Chen2023}
\BIBentryALTinterwordspacing
J.~Chen, R.~G. Edwards, and W.~Mao, ``Graph contractions for calculating correlation functions in lattice qcd,'' in \emph{Proceedings of the Platform for Advanced Scientific Computing Conference}, ser. PASC '23.\hskip 1em plus 0.5em minus 0.4em\relax New York, NY, USA: Association for Computing Machinery, 2023. [Online]. Available: \url{https://doi.org/10.1145/3592979.3593409}
\BIBentrySTDinterwordspacing

\bibitem{Wang2022a}
Q.~Wang, B.~Ren, J.~Chen, and R.~G. Edwards, ``Micco: An enhanced multi-gpu scheduling framework for many-body correlation functions,'' in \emph{2022 IEEE International Parallel and Distributed Processing Symposium (IPDPS)}, 2022, pp. 135--145.

\bibitem{Wang2022b}
\BIBentryALTinterwordspacing
Q.~Wang, Z.~Peng, B.~Ren, J.~Chen, and R.~G. Edwards, ``Memhc: An optimized gpu memory management framework for accelerating many-body correlation,'' \emph{ACM Trans. Archit. Code Optim.}, vol.~19, no.~2, Mar. 2022. [Online]. Available: \url{https://doi.org/10.1145/3506705}
\BIBentrySTDinterwordspacing

\bibitem{Sethi1973}
\BIBentryALTinterwordspacing
R.~Sethi, ``Complete register allocation problems,'' ser. STOC '73.\hskip 1em plus 0.5em minus 0.4em\relax New York, NY, USA: Association for Computing Machinery, 1973, p. 182–195. [Online]. Available: \url{https://doi.org/10.1145/800125.804049}
\BIBentrySTDinterwordspacing

\bibitem{Gilbert1979}
\BIBentryALTinterwordspacing
J.~R. Gilbert, T.~Lengauer, and R.~E. Tarjan, ``The pebbling problem is complete in polynomial space,'' in \emph{Proceedings of the Eleventh Annual ACM Symposium on Theory of Computing}, ser. STOC '79.\hskip 1em plus 0.5em minus 0.4em\relax New York, NY, USA: Association for Computing Machinery, 1979, p. 237–248. [Online]. Available: \url{https://doi.org/10.1145/800135.804418}
\BIBentrySTDinterwordspacing

\bibitem{Boas1979}
P.~van Emde~Boas and J.~van Leeuwen, ``Move rules and trade-offs in the pebble game,'' in \emph{Theoretical Computer Science 4th GI Conference}, K.~Weihrauch, Ed.\hskip 1em plus 0.5em minus 0.4em\relax Berlin, Heidelberg: Springer Berlin Heidelberg, 1979, pp. 101--112.

\bibitem{JiaWei1981}
\BIBentryALTinterwordspacing
H.~Jia-Wei and H.~T. Kung, ``I/o complexity: The red-blue pebble game,'' in \emph{Proceedings of the Thirteenth Annual ACM Symposium on Theory of Computing}, ser. STOC '81.\hskip 1em plus 0.5em minus 0.4em\relax New York, NY, USA: Association for Computing Machinery, 1981, p. 326–333. [Online]. Available: \url{https://doi.org/10.1145/800076.802486}
\BIBentrySTDinterwordspacing

\bibitem{Chung2011}
\BIBentryALTinterwordspacing
C.-C. Lam, T.~Rauber, G.~Baumgartner, D.~Cociorva, and P.~Sadayappan, ``Memory-optimal evaluation of expression trees involving large objects,'' \emph{Computer Languages, Systems and Structures}, vol.~37, no.~2, pp. 63--75, 2011. [Online]. Available: \url{https://www.sciencedirect.com/science/article/pii/S1477842410000278}
\BIBentrySTDinterwordspacing

\bibitem{Liu1987}
\BIBentryALTinterwordspacing
J.~W.~H. Liu, ``An application of generalized tree pebbling to sparse matrix factorization,'' \emph{SIAM Journal on Algebraic Discrete Methods}, vol.~8, no.~3, pp. 375--395, 1987. [Online]. Available: \url{https://doi.org/10.1137/0608031}
\BIBentrySTDinterwordspacing

\bibitem{Agullo2016}
\BIBentryALTinterwordspacing
E.~Agullo, P.~R. Amestoy, A.~Buttari, A.~Guermouche, J.-Y. L'Excellent, and F.-H. Rouet, ``Robust memory-aware mappings for parallel multifrontal factorizations,'' \emph{SIAM Journal on Scientific Computing}, vol.~38, no.~3, pp. C256--C279, 2016. [Online]. Available: \url{https://doi.org/10.1137/130938505}
\BIBentrySTDinterwordspacing

\bibitem{Dubois2015}
\BIBentryALTinterwordspacing
L.~Eyraud-Dubois, L.~Marchal, O.~Sinnen, and F.~Vivien, ``Parallel scheduling of task trees with limited memory,'' vol.~2, no.~2, Jun. 2015. [Online]. Available: \url{https://doi.org/10.1145/2779052}
\BIBentrySTDinterwordspacing

\bibitem{Fradet2023}
\BIBentryALTinterwordspacing
P.~Fradet, A.~Girault, and A.~Honorat, ``Sequential scheduling of dataflow graphs for memory peak minimization,'' in \emph{Proceedings of the 24th ACM SIGPLAN/SIGBED International Conference on Languages, Compilers, and Tools for Embedded Systems}, ser. LCTES 2023.\hskip 1em plus 0.5em minus 0.4em\relax New York, NY, USA: Association for Computing Machinery, 2023, p. 76–86. [Online]. Available: \url{https://doi.org/10.1145/3589610.3596280}
\BIBentrySTDinterwordspacing

\bibitem{Kayaaslan2018}
\BIBentryALTinterwordspacing
E.~Kayaaslan, T.~Lambert, L.~Marchal, and B.~Uçar, ``Scheduling series-parallel task graphs to minimize peak memory,'' \emph{Theoretical Computer Science}, vol. 707, pp. 1--23, 2018. [Online]. Available: \url{https://www.sciencedirect.com/science/article/pii/S0304397517307053}
\BIBentrySTDinterwordspacing

\bibitem{Marchal2019}
\BIBentryALTinterwordspacing
L.~Marchal, B.~Simon, and F.~Vivien, ``Limiting the memory footprint when dynamically scheduling dags on shared-memory platforms,'' \emph{Journal of Parallel and Distributed Computing}, vol. 128, pp. 30--42, 2019. [Online]. Available: \url{https://www.sciencedirect.com/science/article/pii/S0743731518305112}
\BIBentrySTDinterwordspacing

\end{thebibliography}

\end{document}